\documentclass{aa}  
\usepackage{natbib}
\usepackage{graphicx}
\usepackage{txfonts}
\begin{document} 

\title{The Near-Infrared Spectrograph (NIRSpec) on the James Webb Space Telescope}
\subtitle{II. Multi-object spectroscopy (MOS)}

\author{P. Ferruit \inst{1}
\and  P. Jakobsen \inst{2}
\and  G. Giardino   \inst{3}
\and  T. Rawle          \inst{4}
\and  C. Alves de Oliveira \inst{1}
\and  S. Arribas   \inst{5}
\and  T. L. Beck         \inst{6}
\and  S. Birkmann    \inst{4}
\and  T. B\"{o}ker   \inst{4}
\and  A. J. Bunker         \inst{8}
\and  S. Charlot \inst{9}
\and  G. de Marchi     \inst{10}
\and  M. Franx        \inst{11}
\and  A. Henry   \inst{6}
\and  D. Karakla \inst{6}
\and  S. A. Kassin \inst{6},\inst{7}
\and  N. Kumari      \inst{12}
\and  M. L\'{o}pez-Caniego \inst{13}
\and  N. L\"{u}tzgendorf \inst{4}
\and  R. Maiolino   \inst{14}
\and  E. Manjavacas   \inst{12}
\and  A. Marston       \inst{1}
\and  S. H. Moseley     \inst{15}
\and  J. Muzerolle    \inst{6}
\and  N. Pirzkal    \inst{12}
\and  B. Rauscher    \inst{16}
\and  H.-W. Rix    \inst{17}
\and  E. Sabbi        \inst{6}
\and  M. Sirianni     \inst{4}
\and  M. te Plate   \inst{4}
\and  J. Valenti       \inst{6}
\and  C. J. Willott   \inst{18}
\and  P. Zeidler      \inst{12}}
\institute{ 
%1
European Space Agency, European Space Astronomy Centre, Madrid, Spain \and
%2
Cosmic Dawn Center, Niels Bohr Institute, University of Copenhagen, Denmark \and
%3
ATG Europe for the European Space Agency, European Space Research and Technology Centre, Noordwijk, The Netherlands \and
%4
European Space Agency, Space Telescope Science Institute, Baltimore, Maryland, USA \and
%5
Centro de Astrobiolog\'ia, (CAB, CSIC--INTA), Departamento de Astrof\'\i sica, Madrid, Spain \and
%6
Space Telescope Science Institute, Baltimore, Maryland, USA \and
%7
Department of Physics and Astronomy, Johns Hopkins University,  Baltimore, Maryland, USA \and
%8
Department of Physics, University of Oxford, United Kingdom \and
%9
Sorbonne Universit\'{e}, CNRS, UMR 7095, Institut d’Astrophysique de Paris, France \and
%10
European Space Agency, European Space Research and Technology Centre, Noordwijk, The Netherlands \and
%11
Leiden Observatory, Leiden University, The Netherlands \and
%12
AURA for the European Space Agency, Space Telescope Science Institute, Baltimore, Maryland, USA \and
%13
Aurora Technology for the European Space Agency, European Space Astronomy Centre, Madrid, Spain \and
%14
Kavli Institute for Cosmology, University of Cambridge, United Kingdom \and
%15
Quantum Circuits, Inc., New Haven, Connecticut, USA \and
%16
NASA Goddard Space Flight Center, Greenbelt, Maryland, USA \and
%17
Max-Planck Institute for Astronomy, Heidelberg, Germany \and
%18
NRC Herzberg, National Research Council, Victoria, British Columbia, Canada 
}

\date{Received 16 November 2021; accepted 27 January 2022}

% \abstract{}{}{}{}{} 
% 5 {} token are mandatory
 
\abstract{ 
We provide an overview of the capabilities and performance of the  Near-Infrared Spectrograph (NIRSpec) on the James Webb Space Telescope (JWST) when used in its  multi-object spectroscopy (MOS) mode employing a novel Micro Shutter Array (MSA) slit device. The MSA consists of four separate 98\arcsec\ $\times$ 91\arcsec \ quadrants each containing $365\times171$ individually addressable shutters whose open areas on the sky measure 0.20\arcsec\ $\times$ 0.46\arcsec\ on a 0.27\arcsec\ $\times$ 0.53\arcsec\ pitch. This is the first time that a configurable multi-object spectrograph has been available on a space mission.
The levels of multiplexing achievable with NIRSpec MOS mode are quantified and we show that NIRSpec will be able to observe typically fifty to two hundred objects simultaneously with the pattern of close to a quarter of a million shutters provided by the MSA. This pattern is fixed and regular, and we identify the specific constraints that it yields for NIRSpec observation planning. In particular, the roll angle at which a given NIRSpec MSA observation will be executed will, in most cases, not be known before the observation is actually scheduled. As a consequence, NIRSpec users planning MOS mode observations cannot at the proposal stage know precisely which subset of their intended targets will be observable, and will therefore need to intentionally oversize their submitted target catalogues accordingly.
We also present the data processing and calibration steps planned for the NIRSpec MOS data. The significant variation in size of the mostly diffraction-limited instrument point spread function over the large wavelength range of 0.6-5.3~$\mu$m covered by the instrument, combined with the fact that most targets observed with the MSA cannot be expected to be perfectly centred within their respective slits, makes the  spectrophotometric and wavelength calibration of the obtained spectra particularly complex. This is reflected by the inclusion of specific steps such as the wavelength zero-point correction nd the relative path loss correction in the NIRSpec data processing and calibration flow. The processing of spectra of morphologically extended targets will require additional attention and development. These challenges notwithstanding, the sensitivity and multiplexing capabilities anticipated of NIRSpec in MOS mode are unprecedented, and should enable significant progress to be made in addressing a wide range of outstanding astrophysical problems.
}


\maketitle
   
\titlerunning{NIRSpec on the James Webb Space Telescope II}
\authorrunning{P. Ferruit et al.}
 
%________________________________________________________________
\section{Introduction}

The Near-Infrared Spectrograph (NIRSpec) \citep[Paper~I]{Jak21} is one of four instruments on the James Webb Space Telescope (JWST) \citep{Gardner2006}, scheduled for launch in late 2021. NIRSpec was designed and built by the European Space Agency (ESA) with Airbus Defense and Space Germany as the prime contractor. The instrument includes contributions by the National Aeronautics and Space Administration (NASA).

Although NIRSpec was initially designed with multi-object observations of faint high redshift galaxies in mind (see Paper~I), it is actually a versatile instrument that will provide key insights in all four main JWST science themes \citep{Gardner2006}: the end of the dark ages, first light and reionisation; the assembly of galaxies; the birth of stars and protoplanetary systems; and planetary systems and the origins of life. This diversity in science goals is enabled by three complementary observing modes: multi-object spectroscopy (MOS), integral field spectroscopy (IFS) \citep[Paper~III]{Boek21}, and fixed slit spectroscopy (FS) \citep[see][Paper~IV for its application to the study of exoplanets]{Birk21}.

In this paper, we focus on NIRSpec's MOS capability. In MOS mode, NIRSpec can obtain spectra of typically fifty to two hundred astronomical targets simultaneously, over a nine square arcminute field of view, at wavelengths between 0.6 and 5.3~$\mu$m, and at spectral resolving powers ranging from several tens to more than three thousand. The primary scientific driver for the MOS mode was the study of the formation and evolution of galaxies through spectroscopic surveys, although such a powerful capability will have numerous applications in other fields, such as the study of pre-main sequence stars or of brown dwarfs \citep[e.g.][]{Birkmann2017}.

In Sect.~\ref{SectionDesignCapabilities} we present the design and capabilities of NIRSpec in MOS mode. Section~\ref{SectionPerformances} describes in some detail the  multiplexing optimisation strategy developed for the mode (Sect.~\ref{Sec:multi}), the anticipated performance, and some of the particular issues faced in spectrophotometrically calibrating  the data obtained (Sect.~\ref{sec:slit_tran} \& \ref{sec:contamination}). Section~\ref{SectionObservations} summarises the operation of the MOS mode, including exposure preparation (Sect.~\ref{subsec:prep}), target acquisition (Sect.~\ref{subsec:ta}), and post-observation data processing (Sect.~\ref{subsec:mos_proc}). Section \ref{SectionSimulation} presents an example of simulated MOS observations, while Sect.~\ref{SectionConclusion} offers concluding remarks.

%__________________________________________________________________
\section{The design and capabilities of the MOS mode}
\label{SectionDesignCapabilities}

\subsection{Spectroscopy with NIRSpec}

NIRSpec can perform spectroscopy over the near-infrared wavelength range 0.6--5.3~\!$\mu$m. A choice of dispersive elements offers three spectral resolutions: a double-pass prism covering the full wavelength range at $R\!\simeq\!100$, and two sets of three diffraction gratings, giving $R \simeq 1000$ or $R\!\simeq\!2700$ respectively within three overlapping bands. Four long-pass interference filters (labelled with the suffix LP) are employed in conjunction with the gratings for order separation, resulting in eight baseline combinations. An additional combination sees the double-pass prism used together with an uncoated CaF$_2$ filter (labelled CLEAR). See Table \ref{tab:modes} for the list of baseline configurations, and Paper~I for more details on these optical elements.

\begin{figure}
\centering
\includegraphics[width=\hsize]{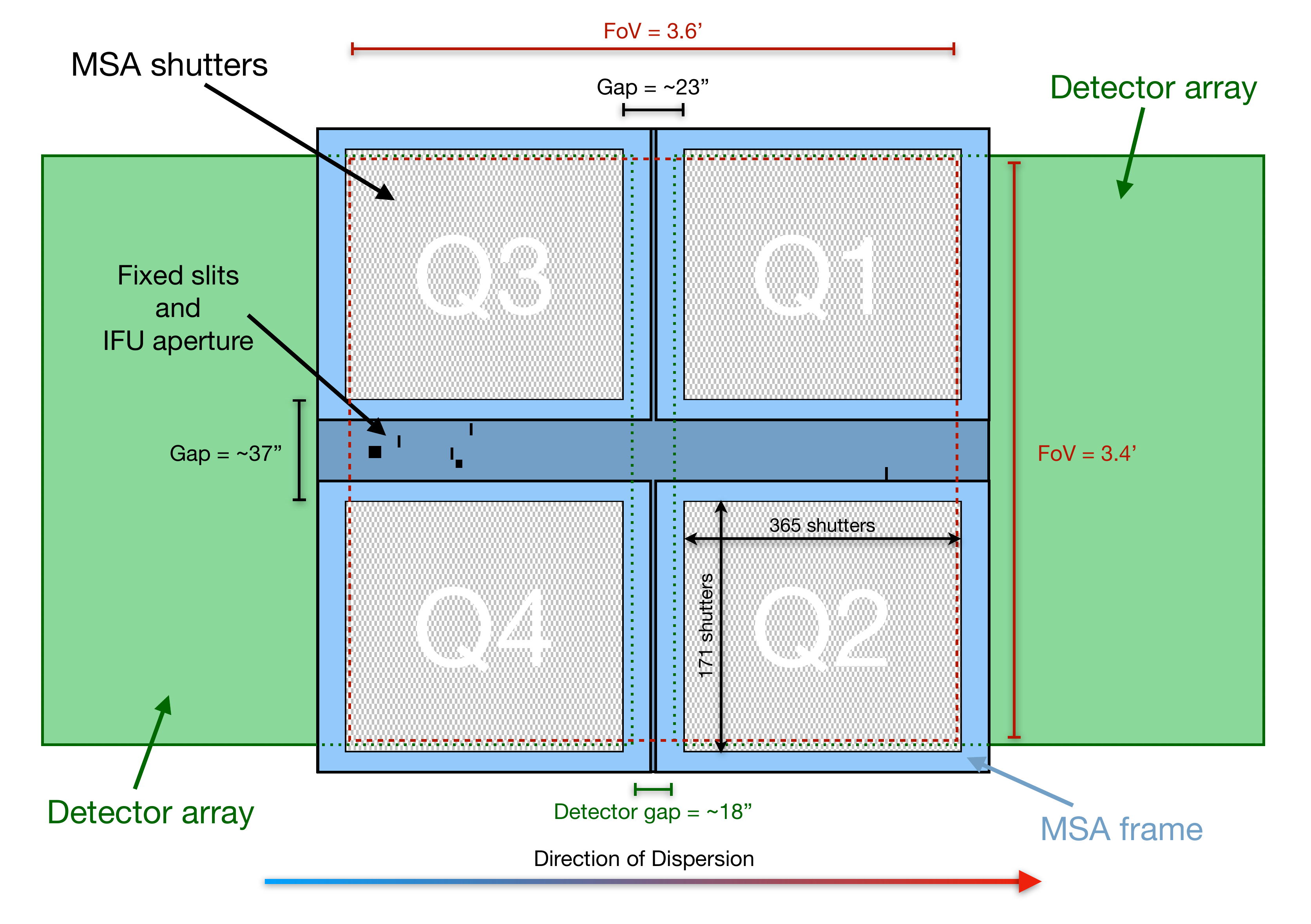}
\caption{Geometry of the NIRSpec aperture plane, including the four MSA quadrants (frame in blue, shutters as a grey checkerboard), the fixed slits and IFU aperture (all in black). Each quadrant, labelled Q1 through Q4, contains 365$\times$171 shutters. The projected locations of the detector arrays are shown in green, with green dotted lines indicating their extent and hence where the gap between detectors is hidden in the figure by the MSA. The red dashed rectangle marks the $\sim$3\farcm 6$\times$3\farcm 4 on-sky field-of-view, constrained by a field stop at the entrance of the instrument.}
\label{fig:slit_plane}
\end{figure}

The NIRSpec slit plane, as shown in Fig.~\ref{fig:slit_plane}, is located at the second image plane of the instrument between the filter wheel and the collimator optics (see Paper~I), and enables three distinct observing modes. Five fixed slits cut into the central plate offer single-object capability, including a large 1\farcs6$\times$1\farcs6 square slit designed for observations of transiting exoplanets (see Paper~IV). A sixth aperture allows light into the Integral Field Unit (IFU), providing spatially resolved spectroscopy over a 3\farcs1$\times$3\farcs2 area on the sky (Paper~III). The majority of the slit plane comprises the four quadrants of the Micro Shutter Array (MSA), enabling multi-object spectroscopy over a $\sim$3\farcm6$\times$3\farcm4 field of view. The remainder of this paper exclusively focuses on the MOS mode of NIRSpec.

   \begin{table}
      \caption{Baseline disperser and filter combinations for NIRSpec. For the grating-based configurations, the quoted wavelength ranges refer to spectral regions free from overlap with any significant second order emission exceeding the 0.1\% level. In the MOS mode, spectra from the high-resolution gratings (italicised, grating names suffixed with H) may suffer significant wavelength cut-off at the red end.}
       \centering
         \label{tab:modes}
         \begin{tabular}{c c c}
            \hline\hline
            \noalign{\smallskip}
           Disperser/Filter&  Wavelength Range&  $R$\\
           &  [$\mu$m]&  \\
            \noalign{\smallskip}
            \hline
            \noalign{\smallskip}
 PRISM/CLEAR&  0.6 -- 5.3& 30 -- 330\\ 
 G140M/F070LP& 0.7 -- 1.3& 500 -- 890\,\,\,\\
 G140M/F100LP& 1.0 -- 1.9& 700 -- 1340\\
 G235M/F170LP& 1.7 -- 3.2& 720 -- 1340\\
 G395M/F290LP& 2.9 -- 5.2& 730 -- 1315\\
 \textit{G140H/F070LP}& \textit{0.7 -- 1.3}& \textit{1320 -- 2395\,\,\,}\\
 \textit{G140H/F100LP}& \textit{1.0 -- 1.9}& \textit{1850 -- 3675\,\,\,}\\
 \textit{G235H/F170LP}& \textit{1.7 -- 3.2}& \textit{1910 -- 3690\,\,\,}\\
 \textit{G395H/F290LP}& \textit{2.9 -- 5.2}& \textit{1930 -- 3615\,\,\,}\\
            \noalign{\smallskip}
            \hline
         \end{tabular}
   \end{table}

\begin{figure}
\centering
\includegraphics[width=0.99\hsize]{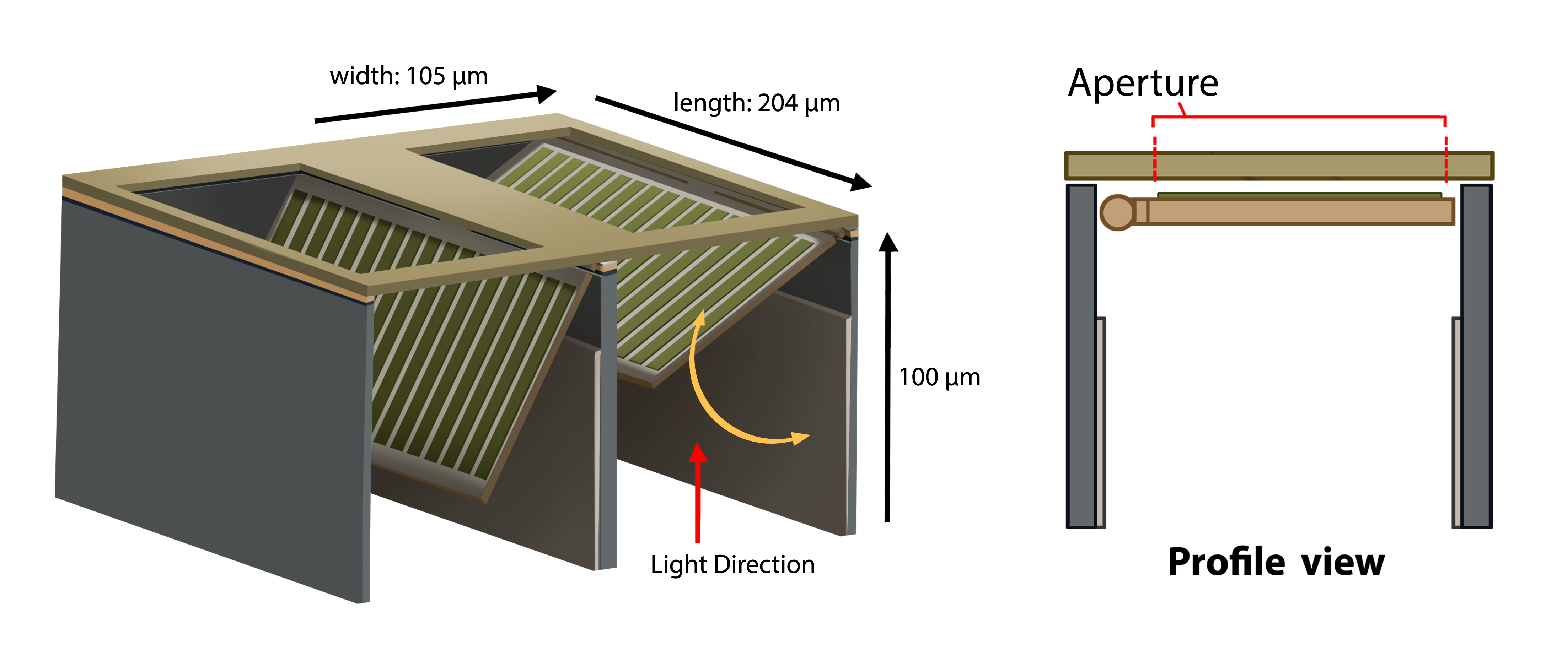}
\caption{Schematic showing the detailed construction of the shutters of the MSA. The width $\times$ length of 105~$\mu$m $\times$ 204~$\mu$m corresponds on average to 268~mas $\times$ 530~mas on the sky. Image credit: Valérian Ferruit.}
\label{fig:shutter_crate}
\end{figure}

\begin{figure*}[p]
\centering
\includegraphics[viewport=20mm 10mm 570mm 278mm,width=0.79\textwidth,clip]{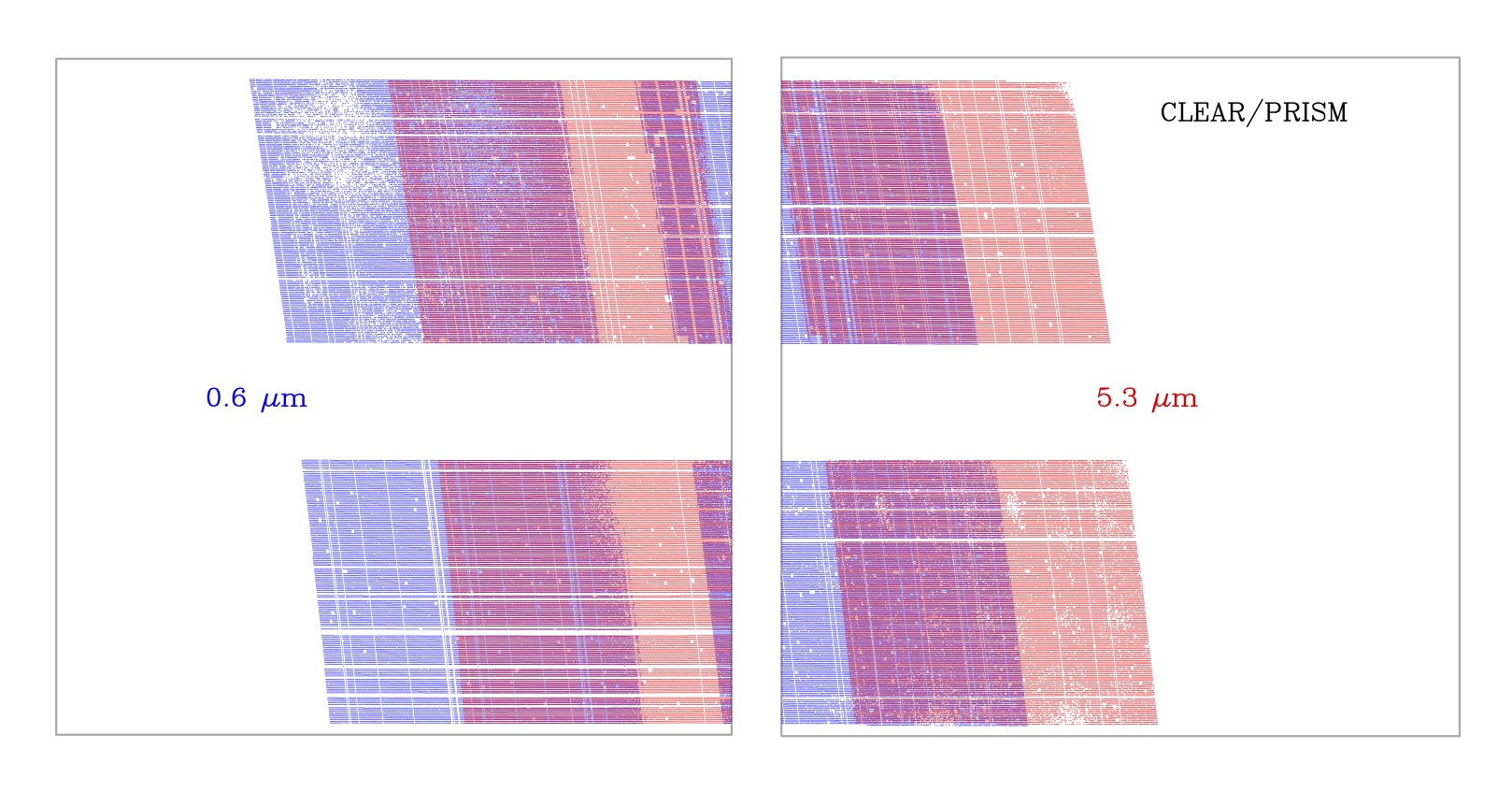}
\includegraphics[viewport=20mm 10mm 570mm 285mm,width=0.79\textwidth,clip]{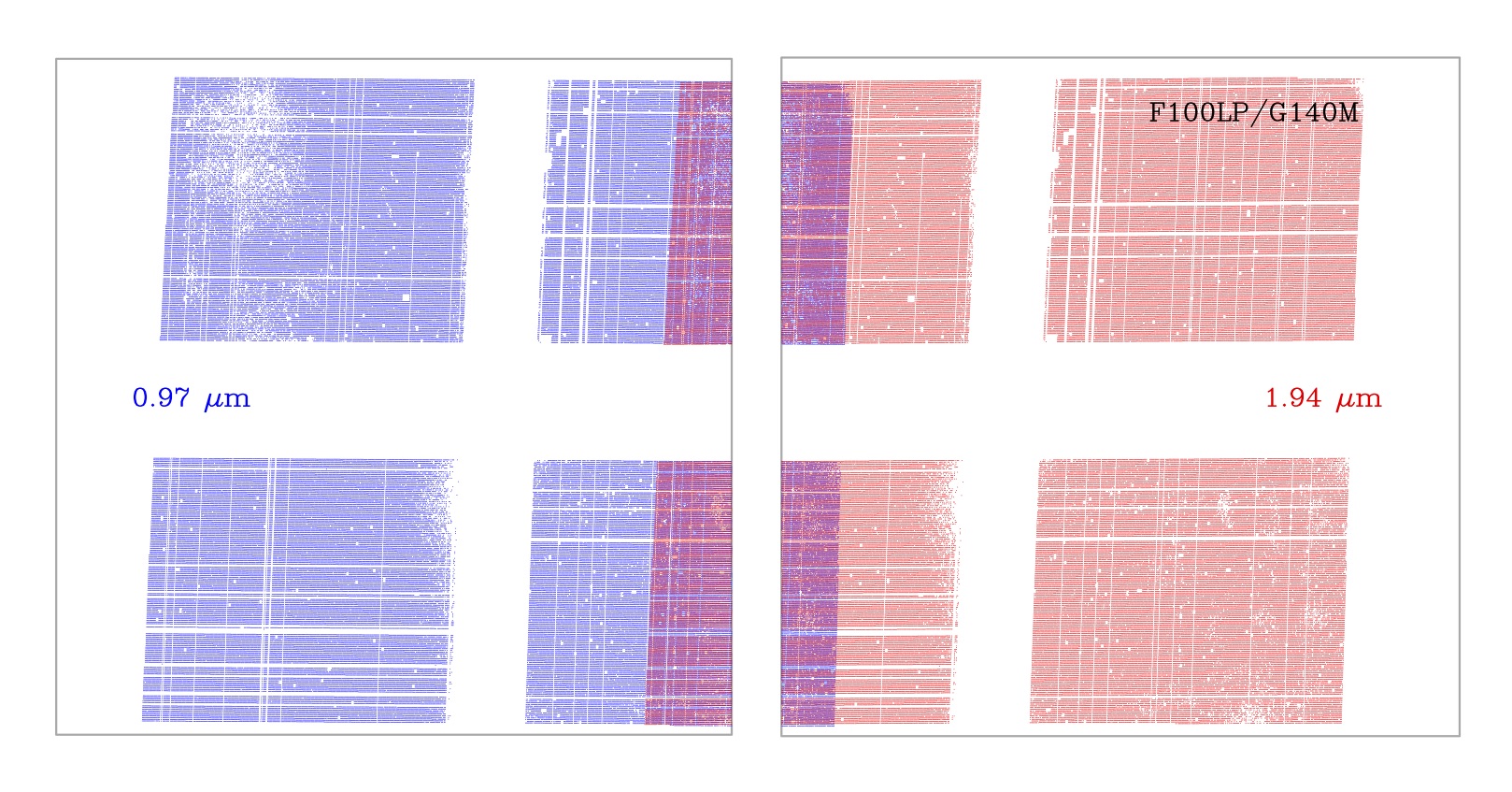}
\includegraphics[viewport=20mm 18mm 570mm 285mm,width=0.79\textwidth,clip]{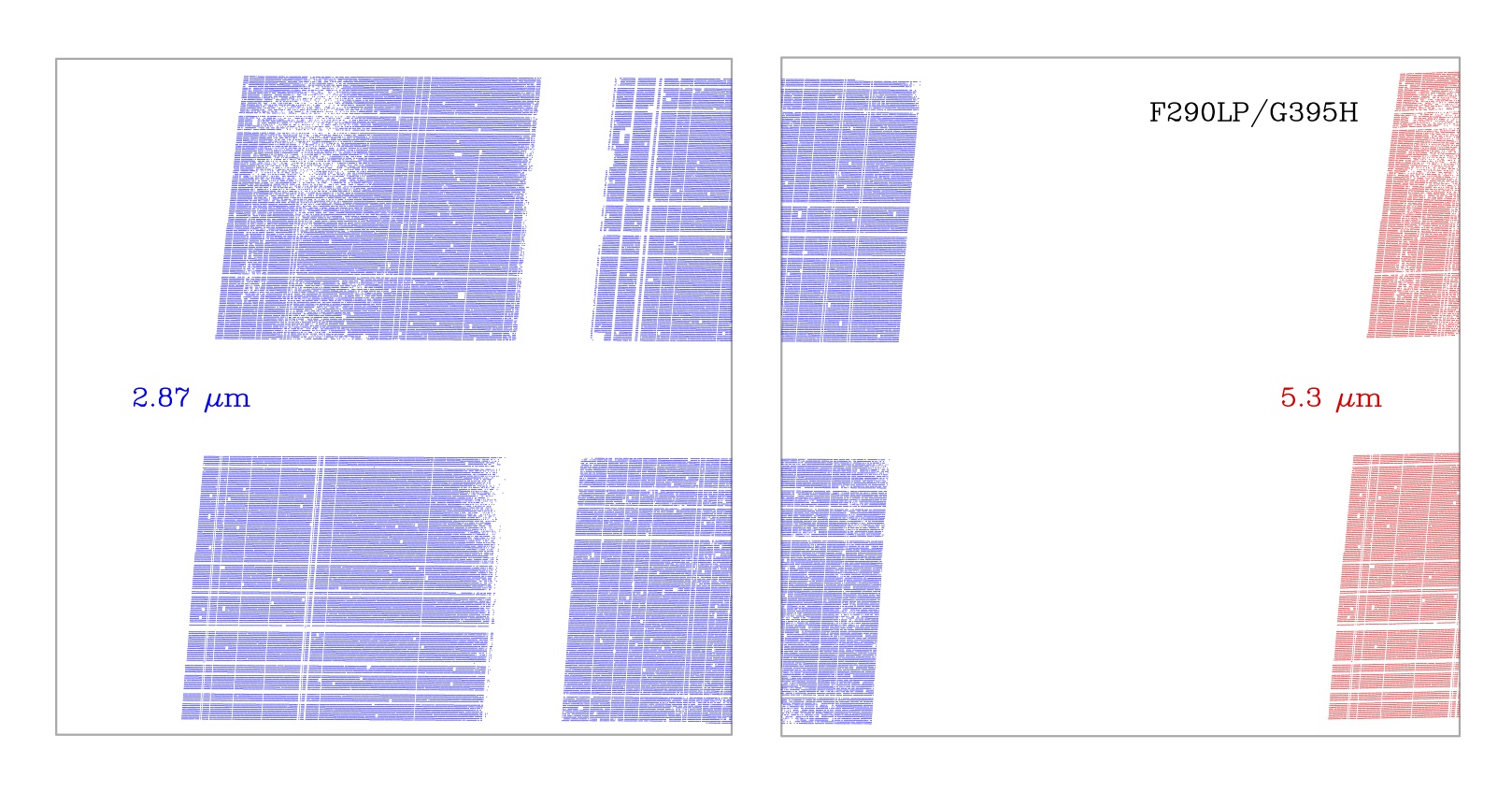}
\caption{Monochromatic footprints of the functional shutters of the MSA, incident on the detector plane (the two outer squares), for the blue and red ends of the nominal wavelength range in three representative filter/disperser combinations (top down, respectively): CLEAR/PRISM (blue $=$ 0.6~$\mu$m, red $=$ 5.3~$\mu$m), F100LP/G140M (0.97 and 1.94~$\mu$m) and F290LP/G395H (2.87 and 5.3~$\mu$m). These plots demonstrate that when using the prism and medium resolution gratings (the example here is representative of all three $R\!\simeq \!1000$ gratings), all the spectral traces connecting the two extreme monochromatic images fall within the outer bounds of the detectors. The plots also reveal the impact of the detector gap, as only spectra from shutters appearing in both colours on a single detector are uninterrupted by the gap (i.e. $\sim$80\% of shutters for the prism, but $<$10\% for G140M). The F290LP/G395H plot also shows that for this, and indeed all  high resolution gratings, only the left-most shutters will have the red end of their spectra falling on the detectors.}
\label{fig:detector_gaps}
\end{figure*}

\subsection{Micro Shutter Array (MSA)}

The MSA was designed and procured by the Goddard Space Flight Center as one of NASA's two hardware contributions to NIRSpec. The assembly provides a fully programmable aperture, consisting of four Micro Electronic Mechanical System (MEMS) array quadrants, each housing 365$\times$171 individually-operable shutters, which yields almost a quarter of a million in total (Fig.~\ref{fig:slit_plane}). For a detailed hardware description, including materials, construction and optimisation of the design, refer to \citet{mose04} and \citet{kuty08}. In this paper, we focus on design details impacting the operation of NIRSpec.

Figure~\ref{fig:shutter_crate} presents the structure of an MSA shutter. Each consists of a door with an embedded electrode (wired to the `171-side' circuit) and etched magnetic strips, hinged at one side to open into a crate-like housing which contains a second electrode on the hinge-side wall (the separate `365-side' circuit). Configuring the MSA involves a coordinated electromagnetic procedure which varies the electrode voltages while sweeping a moving magnet arm back and forth across the shutter arrays. During the outbound journey from the primary park position of the magnet arm near the IFU aperture, in the dispersion direction (towards the shutter hinges), the door and wall electrodes of every shutter are set to opposite potentials, generating an electrostatic force which latches the shutters in place once the magnet pushes them open. After completing this first motion, the magnet arm reverses direction and returns to primary park. Initially, all shutters retain their latching potentials, but as the magnet arm passes over each row in turn, the shutters in that row are individually addressed, holding those that need to remain open (latched), while discharging the voltage for those commanded to close. Shutters with the electrostatic charge removed are gently pulled closed by the passing magnet. In this manner it is possible to configure the MSA into any combination of open and closed shutters.

Each shutter of the MSA has a pitch (distance between the centres of two adjacent apertures) of 105\!~$\mu$m in the axis of dispersion and 204\!~$\mu$m in the perpendicular (spatial) axis. The light-shield frame of the quadrants introduces a $\sim$27\!~$\mu$m bar between neighbouring shutters, yielding an open area per shutter measuring 78\!~$\mu$m $\times$ 178\!~$\mu$m. These physical sizes translate to a predicted average on-sky projection of 268~mas $\times$ 530~mas for the pitch, a $\sim$69~mas bar width and an open shutter size of 199~mas $\times$ 461~mas. The optical field distortion inherent to NIRSpec and the JWST results in plate-scale variations of $1.8$\% in width and $3.6$\% in height over the field of view. Thus the projected shutter pitch ranges between 266~mas and 270~mas in the dispersion direction and between 520~mas and 539~mas in the spatial direction. These numbers will be updated after launch once the telescope is aligned

The micro-shutter width of 78\!~$\mu$m projects geometrically to approximately two pixels on the detectors. This ensures that, by design, NIRSpec MOS line spread function for a fully illuminated spectroscopic aperture is sampled by at least two detector pixels.

Each quadrant has an unvignetted active area of $\sim$96\arcsec \ $\times$ 87\arcsec \, of which $\sim$65\% consists of the shutter apertures themselves. The four quadrants are separated by a cruciform, which introduces a $\sim$23\arcsec \  gap between active areas in the dispersion direction, and a $\sim$37\arcsec \ gap in the cross-dispersion direction. The larger gap in the latter axis includes the $\sim$20\arcsec \ wide plate which is home to the fixed slits and IFU aperture, running through the centre of the slit plane in the dispersion direction (see Fig.~\ref{fig:slit_plane}).

\subsection{MSA spectra on the detectors}

When using the MSA, it is important to understand the connection between the physical location of open shutters and the projected location of the resultant spectral traces on the detector arrays. NIRSpec has two 2048$\times$2048 pixel detectors (see Paper I for further details), abutted in the dispersion direction with a $\sim$18\arcsec \ gap in between. The detectors can be thought of as aligning with the MSA as shown in Fig.~\ref{fig:slit_plane}, such that the gap between the detectors is approximately aligned with the cross-dispersion gap between MSA quadrants. Indeed, in imaging mode (when the grating wheel is rotated to the mirror rather than to a dispersive element), the location of the un-dispersed `footprint' of the MSA on the detectors closely resembles the layout in Fig.~\ref{fig:slit_plane}.

Simulated examples of spectra on the detectors are given in Sect.~\ref{SectionSimulation}: Fig.~\ref{fig:prism_d0} (for the prism) and Fig.~\ref{fig:g235m_d0} (for a medium resolution grating). A few important additional details are worth elaborating. First, the NIRSpec field of view is set by a rectangular field stop measuring $\sim$3\farcm6$\times$3\farcm4 which is part of the instrument Foreoptics and is located in the telescope focal plane, at the entrance of the instrument.  The MSA is, by design, oversized and the outermost approximately ten shutter rows and columns are vignetted and therefore unusable for on-sky targets. The MSA cruciform is an obvious additional area of the full NIRSpec field of view from which MOS targets cannot be observed, as it contains no shutters. This may be employed to a user's advantage, as a bright object in the middle of a field-of-interest can be placed behind the cruciform to avoid contamination (see also Sect.~\ref{sec:contamination}).

Less obvious are the spectral cut-offs, and understanding these is crucial for a successful MOS observing strategy. Although all disperser-filter configurations can be used in any NIRSpec mode, the MOS is primarily designed for use with the $R\!\simeq\!100$ and $R\!\simeq\!1000$ dispersers (highlighted in Table~\ref{tab:modes}). The last three configurations in Table~\ref{tab:modes} employing the $R\!\simeq\!2700$ gratings have spectra that are so long that in MOS mode their red ends are lost for more than half of the shutters (including all of those in Q1 \& Q2). The shorter G140H/LP070 spectra become truncated over a somewhat smaller portion of the MSA. Furthermore, the wavelength coverage of some spectral traces may not be complete due to the physical gap between the two detectors. The wavelength range, if any, affected by this detector gap is dependent on both shutter location and disperser-filter combination: only $\sim$20\% of shutters are affected when using the prism, but the majority are impacted when using a medium resolution grating. When using a high resolution grating, all spectral traces will lose a small range in wavelength due to the detector gap.

Figure~\ref{fig:detector_gaps} helps one to visualise this issue for a representative selection of dispersers by showing the projected monochromatic footprints of the operational shutters of the MSA at the two extremes of the nominal disperser wavelength range. The projection of the MSA shutters at the shortest wavelength of the spectra are shown in blue while the long wavelength edge is in red. The skewed appearance of the single-wavelength MSA image is caused by the out-of-plane incidence of the optical beam on NIRSpec dispersers (Paper~I). The approximate spectral trace arising from any given shutter can be visualised by mentally connecting the red and blue images of the shutter in question (ignoring the optical distortion in the NIRSpec Camera which gives rise to a slight downward curvature to the spectra between their endpoints, see Paper~I). Careful design of an MSA configuration may mitigate the impact of the detector gaps if appropriate shutters can be selected to ensure that scientifically-valuable wavelengths are not missed.

Alternatively, obtaining two exposures with an 18\arcsec \ or more dither along the dispersion direction would ensure acquisition of the complete spectra of selected science sources, as a different wavelength range would fall on the detector gap in each. However, each dither position requires a reconfigured MSA, and the combined effects of the finite MSA quadrant size, optical distortion and the occurrences of failed shutters (see below) mean that two MSA configurations separated by such a large dither are only able to have 30-40\% targets in common, even when targets can be positioned over the full shutter open area. This limits the efficiency of this alternative solution.

\begin{figure}
\centering
\includegraphics[width=0.95\hsize]{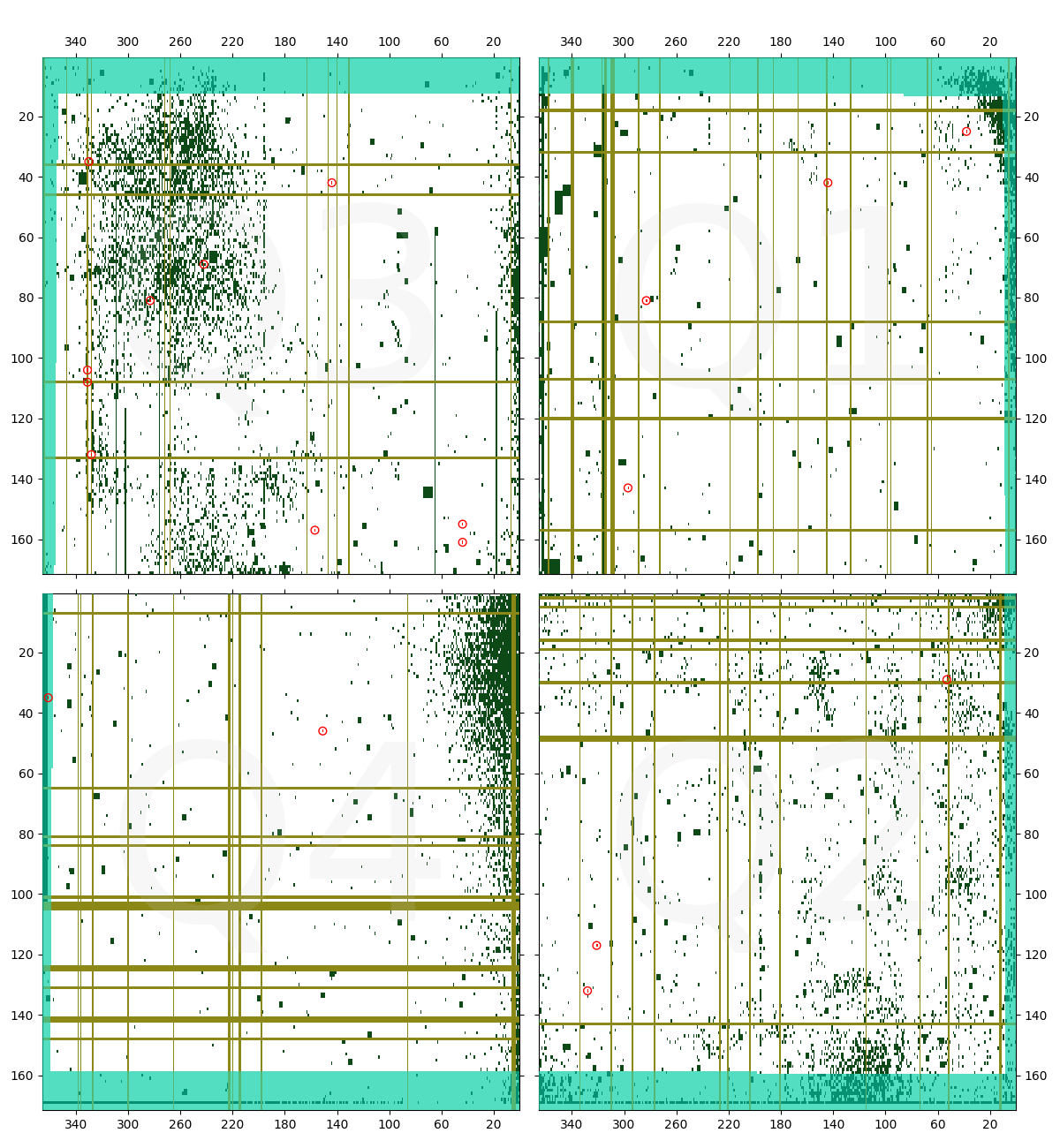}
\caption{Representation of all non-operational shutters as measured at the end of the final cryo-vacuum ground test in 2017 \citep{Rawle+2018}: Failed closed (dark green colour), vignetted shutters (light green colour, and transparent to show failed shutters behind the field stop), and lines and rows of shutters removed from service by the short masks (olive green colour). Failed open shutters are highlighted by red circles. Quadrant orientation is the same as in Fig.~\ref{fig:slit_plane}.}
\label{fig:operability_map}
\end{figure}

\subsection{Shutter operability}

Every NIRSpec exposure employing the MSA can command open a different combination of the nearly quarter of a million shutters by appending an MSA configuration file to the up-linked observation set-up. Configurations for scientific exposures will generally open between a few tens and a few hundred shutters, to ensure optimal packing of the resultant spectra with no overlap on the detector array (see Sect.~\ref{Sec:multi}).

In order to be able to individually command each shutter, the MSA contains very intricate and densely-packed control circuitry, susceptible to low-level particulate contamination causing occasional electrical shorts. Shorts can cause a high current, unpredictable shutter behaviour and/or a thermal glow detectable by the NIRSpec sensors as parasitic light. To prevent these consequences, known shorted rows or columns are placed into a special mask loaded onto the microshutter control electronics (MCE), which effectively removes them from the active circuitry, overriding any subsequent commanding to keep the shutters closed and ultimately protecting the hardware.

Shutters can also occasionally warp, which causes them to stick in either the open or closed position. Failed open (FO) shutters do not close when commanded, and cause an unwanted spectral trace on the detector. Identification of FO shutters is critical so users can ensure that science spectra do not coincide with the contamination. Thankfully, so far, the FO shutter population is small and stable, with fewer than 20 apparent during ground testing (\citealt{Rawle+2018}; highlighted in red in Fig.~\ref{fig:operability_map}), some of which are actually shutterless apertures since failure early in quadrant component testing. If a significant increase of the FO shutter population is observed in orbit, a contingency annealing procedure will be exercised to try to heal the new FO shutters. This procedure heats the MSA quadrants for several minutes, helping shutters unstick, before letting them return to the nominal operating temperature. Annealing was executed in ground testing and appeared to heal at least one recently failed open shutter, but we note that the instrument was essentially annealed between cryo-testing and flight, so any pre-launch failed open shutters are unlikely to respond further to the tactic. In  flight, the procedure will be a last resort in case of an unusable build-up of stuck open shutters, as the heating process is highly disruptive, likely requiring a subsequent full re-calibration of the instrument model.

Failed closed (FC) shutters are those that either remain closed when initially opening the array, or fail to hold open when commanded to do so. A few tens of thousands of shutters are FC during diagnostic configurations (\citealt{Rawle+2018}; coloured dark green in Fig.~\ref{fig:operability_map}). Combined with the shutters placed into the short masks, $\sim$14\% of all shutters are not operational. Fortunately, their impact is much less critical than for FO shutters, and merely precludes a particular part of the array from being used as an open slit. As long as the percentage of these shutters does not become too large, the multiplexing capability of NIRSpec is not severely disrupted. Furthermore, the complexity of the commanded configuration impacts the number of FC shutters, as the drive electronics are forced to switch more frequently causing intermittent failures. This implies that some FC shutters are transient and may occur at any time. For a nominal science configuration containing a few hundred commanded-open shutters, we only expect $\sim$0--1 of them to be affected by a statistical failure and behave as a transient failed closed.

The degree to which the pattern of operational shutters will evolve, both during launch and with use in flight, is poorly known. It is sobering to consider that the MSA was last used and inspected in the summer of 2017 during the observatory level cryo-vacuum test campaign at Johnson Space Center \citep{Rawle+2018}, and will only be activated again after launch, once the instrument is at its operational cryogenic temperature. Thereafter, the state of the MSA will be monitored regularly using uniform illumination exposures obtained with NIRSpec's internal calibration lamps (cf. Fig.~\ref{fig:msa_contrast}).

When developing NIRSpec observing programmes, one must find optimal MSA configurations that include the maximum number of catalogue targets, accounting for the MSA geometry and the latest knowledge of the inoperable shutters (FC and FO shutters). The next section delves into the strategies and limitations of this multiplexing optimisation.

%__________________________________________________________________
\section{MOS mode performance}
\label{SectionPerformances}

\subsection{Observing approach and multiplexing}
\label{Sec:multi}

The NIRSpec MSA makes for a rather unconventional slit device for a multi-object spectrograph, in that its regular grid of shutters projects to a static semi-regular pattern of available slits on the sky that can only be matched to a collection of targets through shifting the MSA projection in bulk by slewing the telescope. The four spaced MSA quadrants, together with the not insignificant number of non-operational shutters that they each contain, makes the pattern of available shutters quite complex and non-contiguous in shape (cf. Figs.~\ref{fig:slit_plane} and \ref{fig:operability_map}). A further complication is that distortion in the NIRSpec collimator optics, and to a lesser extent small misalignment errors between the MSA quadrants, as well as residual clocking errors in the NIRSpec dispersers, cause the spectral traces from different shutters to overlap on the detector array in a complicated manner that depends on the disperser employed.
The challenge NIRSpec users face in designing an MSA observation is to determine the best combination of telescope pointing and pattern of shutters to be commanded open that provides the optimal set of non-overlapping spectra of the highest priority candidate targets.

To accomplish this complex task, a specific tool called the MSA Planning Tool (MPT) \citep{kara14} has been developed by STScI as part of the STScI Astronomer's Proposal Tool (APT) (see Sect.~\ref{subsec:prep}). NIRSpec MOS mode users have to make use of the MPT, optionally supplemented by the so-called eMPT software suite developed by the NIRSpec Instrument Team \citep{pbs+2020}, a version of which will be released publicly in 2022. Both systems employ as their backbone the NIRSpec Instrument Model \citep{dorn16,birk16,glf+16}, on which they rely to accurately project targets from the sky onto the MSA, trace spectra from individual shutters on to the NIRSpec detector array, as well as provide the detailed up-to-date inventory of operational shutters on the MSA.  The MPT and eMPT  mainly differ in the algorithms they each employ to arrive at the optimal MSA configuration solution in a given situation. The level of multiplexing achievable with the two systems are comparable. To quantify the multiplexing, we briefly describe a simple statistical model of the steps performed by the eMPT. Further details will be presented in \citet{pbs+2020}.

Consistent with the standard background-subtraction strategy of NIRSpec MOS mode, all targets are assumed to be observed in three shutter tall slitlets, and that the telescope is nodded by one vertical shutter pitch between each sub-exposure such that all objects being observed are shifted in sequence  between each of the three shutters making up each slitlet. These nods serve to shift all spectra on the detector array by approximately $\pm5$ pixels in the spatial direction between sub-exposures, with the aim of smoothing out any detector blemishes and residual flat-fielding errors when the spectra are extracted and combined during the data reduction (Sect.~\ref{subsec:mos_proc}). As the MSA is shifted on the sky such that the targets being observed are first placed in the central shutters of their slitlets, then in the topmost shutters, and finally in the lower shutters, each nodded slitlet in total samples five different shutter open area projections on the sky along the spatial direction  (i.e. the three shutter views of the sky from the first central exposure, plus one additional flanking view from each of the two nods).  In the case of compact targets whose flux is confined to a single shutter, the nodded three-shutter approach allows `same shutter' local sky background subtraction to be performed in the temporal domain without needing to reconfigure the MSA. Accurate background subtraction in the case of extended objects whose flux extends over the majority of shutters sampled by the nodded slitlet will have to rely on the sky background spectrum being measured elsewhere on the MSA. This approach is referred to as Master Background Subtraction (Paper~I).

The planning of a NIRSpec MSA observation starts with a prioritised input catalogue listing the accurate coordinates of the targets to be observed, together with a placeholder nominal position on the sky towards which the NIRSpec MSA fiducial point -- defined by the centre of the 3\farcm6$\times$3\farcm4 FOV spanned by the unvignetted MSA border -- is to be positioned. As discussed in Sect.~\ref{subsec:prep}, because JWST is severely roll-constrained, it is only possible to know the roll orientation at which any exposure will be carried out once the observation has been scheduled and placed on the detailed observatory timeline. As a consequence, at the proposal stage the user cannot know precisely which subset of catalogue targets it is possible to observe simultaneously in any exposure. To compensate for this uncertainty, the footprint of the NIRSpec input catalogue is allowed to be intentionally oversized (typically extending out to $\simeq\!3$~arcmin radius from the specified nominal pointing), in order to allow the observation to be optimised at the actual roll angle by permitting the final telescope pointing to deviate from the nominal proposed one by a modest amount too small to perturb the observatory scheduling (typically up to $\pm\!30$~arcsec in both RA and Dec). 
 
At a given projection of the MSA on the sky, the first task of the eMPT is to project the targets in the input catalogue onto the NIRSpec slit plane and determine which of these  fall within available shutters of the four quadrants of the MSA. This projection requires accurate knowledge of the combined optical distortion of the NIRSpec Foreoptics and the telescope proper (Paper~I). This sky-to-MSA transformation will only be known to sufficient accuracy after launch once the telescope has been successfully deployed, phased and focused on orbit. The transformation contained in the prelaunch instrument model is a placeholder, derived from ray-tracing the nominal as-designed JWST.

Since the  majority of the input catalogue targets at any orientation of the MSA cannot all be expected to project near any shutter centres, the user needs to specify two dimensionless parameters $0<\theta_x\le0.5$ and $0<\theta_y\le0.5$ corresponding to the fraction of the shutter-to-shutter centre spacing (`pitch') in each direction. $\theta_x$ and $\theta_y$ define the so-called Acceptance Zone determining how far the position of any candidate target (as defined by its catalogue coordinates) is allowed to deviate in either direction from the centre of the shutter that it projects onto. An Acceptance Zone spanning the full shutter open area corresponds to the choice $\theta_x\!=\!0.371$ in the dispersion direction and $\theta_y\!=\!0.436$ in the spatial direction. Numerically smaller values correspond to demanding tighter alignment between the selected target coordinates and their shutter centres, which - depending on the nature and surface brightness distribution of each target being observed - generally leads to more light from each target passing through the open shutter (Sect.~\ref{sec:slit_tran}), albeit at the price of fewer targets being observed overall (unless the target density is extremely high).

Central to the architecture of the eMPT is the so-called Viable Slitlet Map, whose entries code for whether a given MSA shutter is available to serve as the centre of a complete three shutter tall slitlet flanked by operational shutters both above and below it. The Viable Slitlet Map is initialised at the start of each eMPT run for the specified disperser, based on the current inventory of operational shutters, and by masking out any slitlets whose spectra would overlap with the always present single-shutter spectra from the failed open shutters. In the case of the PRISM disperser, slitlets giving rise to spectra that encounter the 2.8~mm gap between the two detector arrays are also masked out (cf. Fig.~\ref{fig:prism_d0}). At the time of writing, the MSA carries a total of $N_\mathrm{S}\!=\!123\,533$ Viable Slitlets in PRISM mode and from $N_\mathrm{S}\!=\!124\,186$ to $124\,210$ Viable slitlets for the three $R\!=\!1000$ grating modes, out of $N_\mathrm{0}\!=\!223\,097$ possible for an ideal fully-functional MSA. It follows that when employing three shutter tall slitlets, the $\simeq$14\% non-operational shutters presently in the pre-flight MSA result in it operating with a total effective active field of view  that is $N_\mathrm{S}/N_\mathrm{0}\!\simeq\!55$\% of that of a hypothetical flawless device. The reason for this large reduction in the active field of view is that a single isolated failed closed shutter precludes not only itself, but also the shutters immediately above and below it, from serving as the central shutter of a fully functioning three shutter tall Viable Slitlet.

At a given pointing and orientation of the MSA on the sky, the first two tasks performed by the eMPT can be readily modelled statistically provided the targets to be observed are assumed to be distributed at random on the sky. If the average surface density of the catalogue targets of interest  is $dn/d\Omega\!=\!\Sigma_\mathrm{T}$ then the number of targets contained within the MSA field of view $n_\mathrm{MSA}$ will be Poisson distributed with mean:
\begin{equation}
\bar{n}_\mathrm{MSA} =  \Omega_\mathrm{MSA} \ \Sigma_\mathrm{T};
\label{eq:nmsa}
\end{equation}
where $\Omega_\mathrm{MSA}\!=\!N_\mathrm{0} \thinspace \Omega_\mathrm{SH}\!=\!9.18$~arcmin$^2$ is the solid angle subtended by the central shutters of the $N_\mathrm{0}\!=\!223\,097$ theoretically possible Viable Slitlets in the four quadrants of the unvignetted  field of view of the MSA, and $\Omega_\mathrm{SH}\!=\!268 \times 529$~mas$^2$=$3.94\times 10^{-5}$~arcmin$^2$ is the mean solid area extended by a single shutter facet. 
The probability that any one of these $n_\mathrm{MSA}$ targets falls inside the Acceptance Zone of a functioning Viable Slitlet is then:
\begin{equation}
p_\mathrm{VS}=\left(\frac{N_\mathrm{S}}{N_\mathrm{0}}\right) \left(\frac{\theta_x}{0.5}\right) \left(\frac{\theta_y}{0.5}\right);
\label{eq:pvs}
\end{equation}
where the first factor is the fraction of available Viable Slitlets on the MSA for the disperser being employed, and the two last factors give the filling factor of the adopted Acceptance Zone with respect to the shutter pitch. 

The next task performed by the eMPT is to exclude from further consideration any 
otherwise valid candidate targets whose nodded slitlets also happen to capture contaminating light from any other catalogue object. The probability that the area sampled by a  given nodded slitlet (equivalent to the area of five open shutters) is free of extraneous contaminating targets beyond the target itself is approximately:
\begin{equation}
p_\mathrm{NC}  \simeq \exp(- 5  \thinspace \Sigma_\mathrm{C} \thinspace  \Omega_\mathrm{SH});
\label{eq:pnc}
\end{equation}
where  $\Sigma_\mathrm{C}$ is the surface density of all known potentially contaminating sources on the sky (i.e. $\Sigma_\mathrm{C} \ge \Sigma_\mathrm{T}$ depending on the completeness of the input catalogue). We note that eq.~(\ref{eq:pnc}) underestimates the contamination in the case of extended targets, whose flux can still extend into the open area of a shutter even when the target photocentre falls well outside the shutter in question. 

Following these two target elimination steps, the number of remaining contamination-free targets located within Acceptance Zones of Viable Slitlets $n_{VS}$ will follow a thinned Poisson distribution with mean:
\begin{equation}
\bar{n}_\mathrm{VS} = p_\mathrm{VS} \ p_\mathrm{NC} \ \bar{n}_\mathrm{MSA} = p_\mathrm{VS} \ p_\mathrm{NC}  \ \Omega_\mathrm{MSA} \ \Sigma_\mathrm{T}.
\label{eq:nvs}
\end{equation}

The next task is to identify the optimal subset  $n_\mathrm{SP}$ of the $n_\mathrm{VS}$ remaining candidate targets whose spectra can be fit onto the $2 \times 2048 \times 2028$ pixel detector array without incurring overlap. The outcomes of this step differ significantly between using the PRISM or one of the $R\!=\!1000$ gratings. As discussed in Paper~I, first order  $R\!=\!1000$ grating spectra are flanked by zero-order images at their blue ends and second-order extensions at their red ends, resulting in their effectively spanning the entire width of the detector. PRISM spectra on the other hand are considerably shorter and possess no higher or lower order images. Up to four PRISM spectra can therefore in principle be placed across the width of the detector array. As a result, typically up to four times more non-overlapping PRISM spectra can be fit on the detector than grating spectra. 

This last down-selection step is carried out in the eMPT employing the Arribas Algorithm. This algorithm takes as its input an arbitrary set of target-containing Viable Slitlet locations on the MSA and determines which of the spectra mutually overlap vertically (as well as horizontally in the case of the PRISM) within the set. The algorithm works by successively identifying and removing the `worst offender' slitlet whose spectrum overlaps with the largest number of other spectra, and repeating the process until none of the remaining spectra mutually overlap. The output of the algorithm is the largest possible subset drawn from the input targets whose spectra can all be accommodated on the detector simultaneously.

\begin{figure}
\includegraphics[width=\hsize]{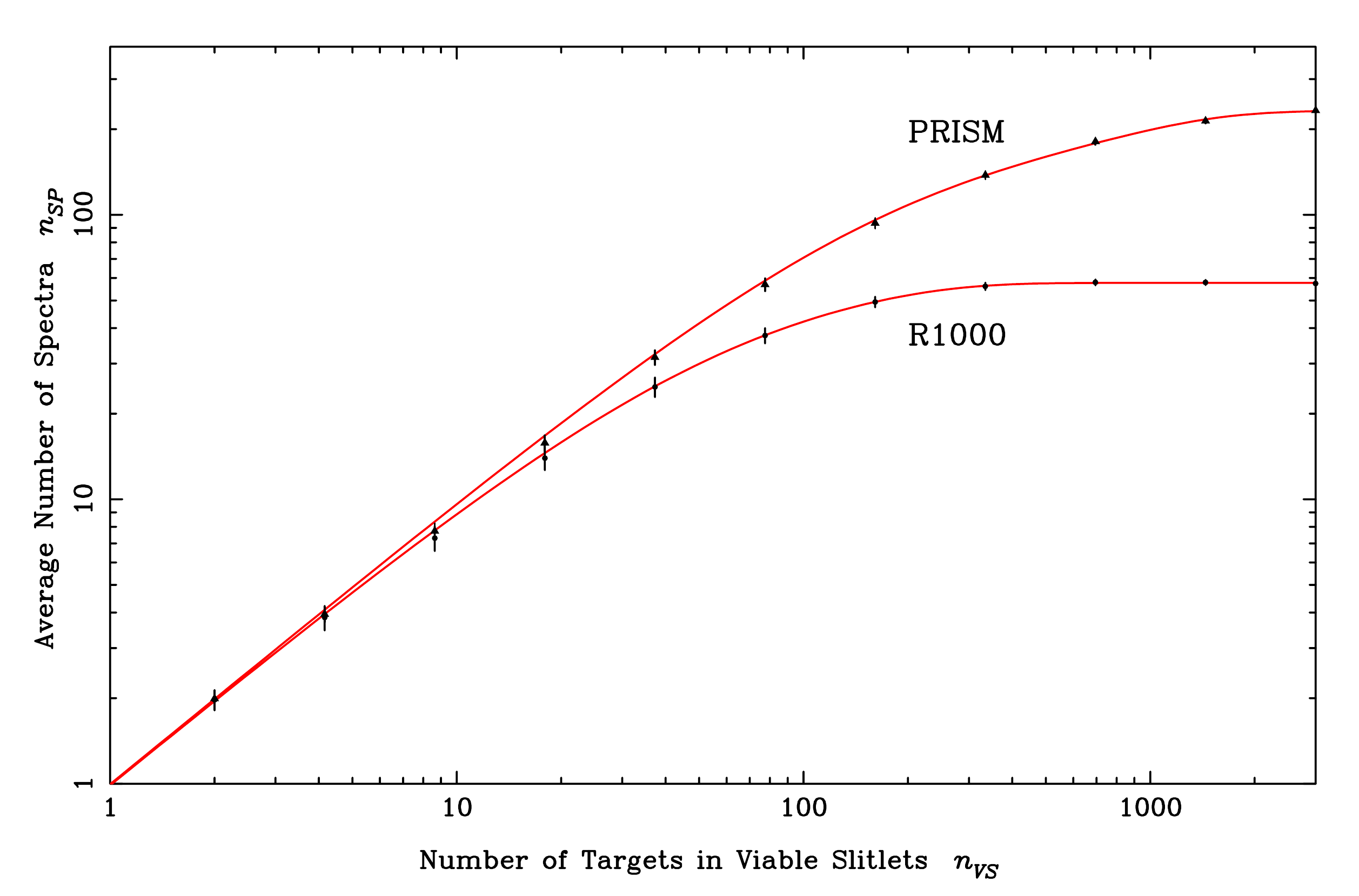}
\caption{Average number of spectra $\bar{n}_\mathrm{SP}$ that can be accommodated on the NIRSpec detector array in MOS mode without overlap  as a function of $n_\mathrm{VS}$, the number of candidate targets located within Viable Slitlets on the MSA. The points show the results of 10\,000 Monte Carlo trials carried out at the shown values of $n_\mathrm{VS}$ for the PRISM and the G235M grating. The continuous curves show the fits to Eq.~(\ref{eq:nsp}).}
\label{fig:multi}
\end{figure}

Due to the complex interplay between the pattern of Viable Slitlets on the MSA and how  their dispersed images project to the NIRSpec detector array, the quantitative performance of the MOS target placement (given the MSA operability) as a function of input target density is best assessed through Monte Carlo simulation. Figure~\ref{fig:multi} plots the mean values of $\bar{n}_\mathrm{SP}$ and the $\pm1\sigma$ scatter in ${n}_\mathrm{SP}$ resulting from 10\,000 trials in which a fixed number $n_\mathrm{VS}$
of Viable Slitlets were picked at random over the MSA and passed to the Arribas Algorithm.

The upper points refer to the PRISM and the lower ones to the G235M grating runs. The results for the other two $R\!=\!1000$ gratings are for practical purposes identical. It is evident that the finite real estate of the detector array is the limiting factor deciding the achievable multiplexing in both grating and PRISM mode. Both multiplexing curves are linear ($\bar{n}_\mathrm{SP} \simeq n_\mathrm{VS}$) at small values of $n_\mathrm{VS}$ -- where spectral collisions are improbable -- and gradually converge to constant values of $\bar{n}_\mathrm{SP}$ for larger values of $n_\mathrm{VS}$ as the detector becomes saturated with spectra. 

The dependency of $\bar{n}_\mathrm{SP}$ on $n_\mathrm{VS}$ can be fit remarkably well by the simple analytic expression:
\begin{equation}
\bar{n}_\mathrm{SP} = m_s \left( 1 -\exp{\left(-{\frac{n_\mathrm{VS} \thinspace \phi(n_\mathrm{VS})}{m_s}}\right)}\right);
\label{eq:nsp}
\end{equation}
where  $m_s$ is the maximum achievable value of $\bar{n}_\mathrm{SP}$, and  $\phi(x)$ is a soft step function:
\begin{equation}
\phi(x) = \alpha_o + (1-\alpha_o)\exp(-\beta_o  x ),
\label{eq:step_function}
\end{equation}
and $\alpha_o$ and $\beta_o$ are fitted constants. The two continuous curves in Fig.~\ref{fig:multi} show the best fits for the PRISM ($m_s\!=\!232.1$, $\alpha_o\!=\!0.427$, $\beta_o\!=\!0.0032$) and G235M grating ($m_s\!=\!57.6$, $\alpha_o\!=\!0.638$, $\beta_o\!=\!0.011$).
These fits, substituting the mean $\bar{n}_\mathrm{VS}$ from eq.~(\ref{eq:nvs}) for $n_\mathrm{VS}$ in eq.~(\ref{eq:nsp}), concisely capture the anticipated ultimate multiplexing capabilities of the pre-launch NIRSpec in MOS mode, and its dependency on the choice of disperser and size of the Acceptance Zone. It is worth noting that since the Arribas Algorithm places the largest possible number of non-overlapping spectra on the detector, the above fits provide strict upper limits on the degree of multiplexing achievable with three shutter tall slitlets employing any alternative target placement scheme.

It is evident that at the target densities $\Sigma_\mathrm{T}\!\ge\!250$~arcmin$^{-2}$ encountered in existing remote galaxy surveys \citep[cf.][]{whit19} and assuming an Acceptance Zone spanning the full shutter open area, the MSA will be partially to fully saturated in $R\!=\!100$ PRISM mode, allowing $\bar{n}_\mathrm{SP}\!\simeq\!150-230$ non-overlapping low resolution galaxy spectra to be observed simultaneously in each exposure. This is in marked contrast to the nearly always saturated $R\!=\!1000$ grating modes which are only capable of capturing $\bar{n}_\mathrm{SP}\!\simeq\!50-60$ non-overlapping intermediate resolution spectra in a single MSA configuration. This significant mismatch between the multiplexing capabilities at $R\!=\!100$ and $R\!=\!1000$ is an issue when attempting to survey a given sample of galaxies at both spectral resolutions. This has led to the approach of carrying out $R\!=\!1000$ exposures in the same MSA configuration optimised for a preceding PRISM observation, thereby accepting a high degree of overlap between grating spectra on the grounds that the higher resolution spectra are primarily concerned with measuring nebular emission lines, which occupy a much smaller filling factor on the detector compared to the underlying continuum spectra (Sect.~\ref{SectionSimulation}).

It is also clear that because of the late assignment of the roll angle, only a small fraction of the potential NIRSpec targets listed in the input catalogue at the proposal stage can be observed with NIRSpec in MOS mode in a single exposure. For example, if the oversized input catalogue extends out to 3~arcmin radius, the $\Omega_\mathrm{SH}\!=\!9.18$~arcmin$^2$ unvignetted MSA field of view will cover only a third of the catalogue targets in any given pointing. Of these targets roughly another third will fall within the all-open Acceptance Zones of Viable Slitlets. Depending on the number of remaining targets, typically yet another third will result in non-overlapping spectra in PRISM mode. In total only a few percent of the targets contained in the oversized input catalogue can be expected to be covered in a single MSA exposure, and seldom more than $\simeq$10\%.

In any actual NIRSpec programme,  not all the targets in the input catalogue will be of equal interest scientifically as implicitly assumed above. In the eMPT the targets in the input catalogue are therefore grouped into a modest number of discrete Priority Classes (typically 5-10) with placements attempted on the MSA in sequence of decreasing priority using the Arribas Algorithm. This results in progressively fewer targets being accepted for the later Priority Classes, and in the resulting MSA configuration containing fewer targets overall than if the Arribas Algorithm were applied in a single pass with all targets pooled together. An example of an eMPT-produced MSA configuration forms the basis for the simulated MSA exposures presented in Sect.~\ref{SectionSimulation}.

\subsection{Sensitivity: The impact of source centring}
\label{sec:slit_tran}

\begin{figure} 
\centering
\includegraphics[width=\hsize]{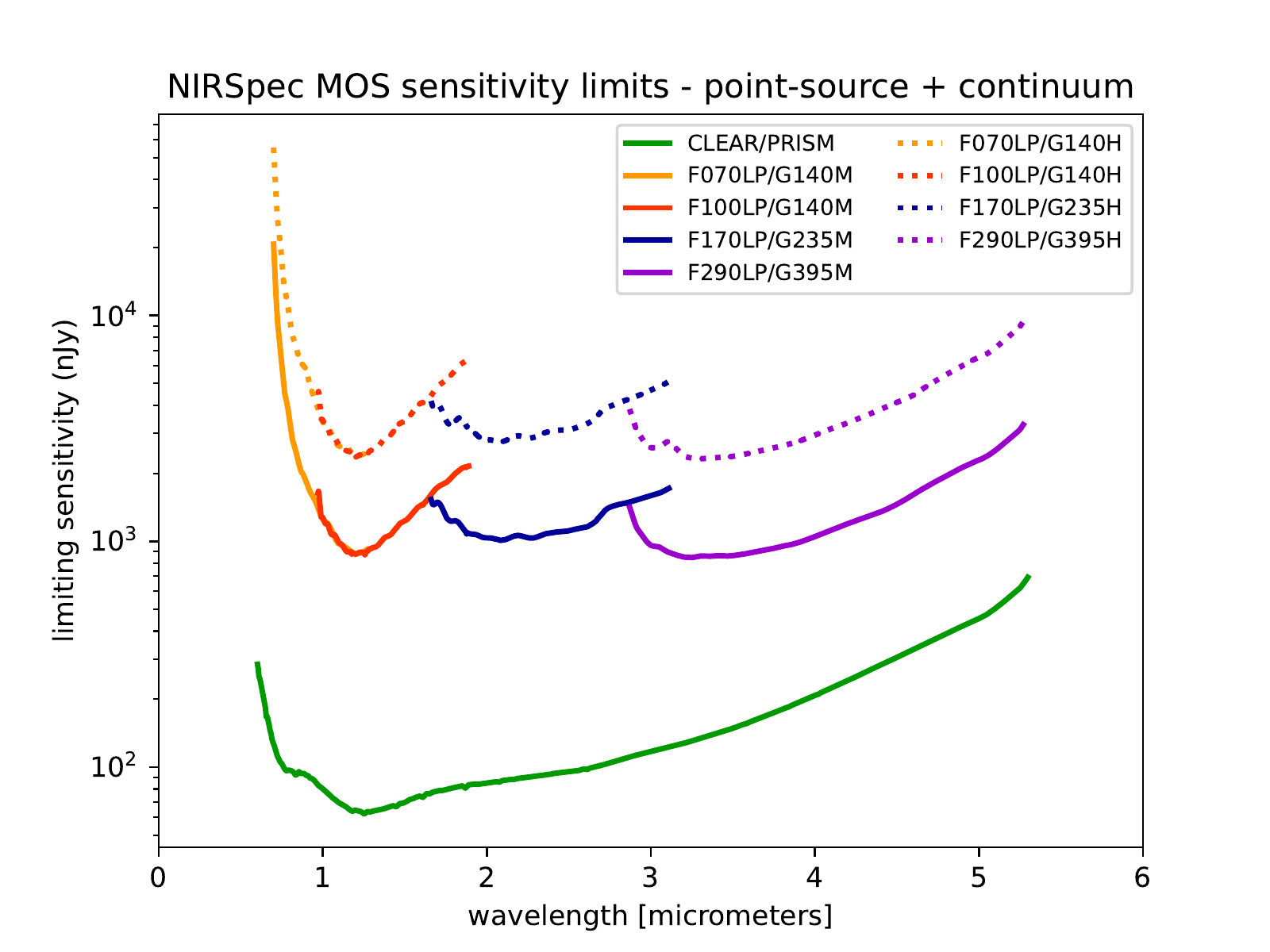}
\caption{NIRSpec sensitivity for a
point source centred in a microshutter aperture, using the 
NRSIRS2 readout mode. The plots show for each disperser the continuum sensitivity as a function of wavelength required to reach $S\!N=10$ per spectral pixel in ten NRSIRS2RAPID exposures with an individual integration time of 1006.7~s (70 groups of 1 frame). The computation was performed using the signal to noise calculation scheme described in Appendix A of Paper I}
\label{fig:mos_sensitivity}
\end{figure}

As discussed in Paper~I, the radiometric sensitivity of NIRSpec in MOS mode is determined by the combination of the instrument throughput, the detector noise for the read-out scheme employed, and the fraction of the total light from the target being observed that passes through the shutters being used, which in turn depends on whether the target is a point source or an extended object, and how the object is positioned within the shutter.  Appendix~A of Paper~I describes in detail the recipe for calculating the limiting sensitivity in MOS mode for two specific benchmark observations used to monitor the anticipated performance of the instrument. Figure~\ref{fig:mos_sensitivity} shows the results of an expanded calculation following the same general recipe, assuming the average noise characteristics of NIRSpec's two detector arrays, and ignoring the impact of cosmic rays hits. For each of NIRSpec's seven dispersers the point source continuum flux required to reach $S\!N\!=\!10$ per single pixel wavelength bin in ten NRSIRS2RAPID exposures, each with an integration time of 1006.7~s (70 groups of 1 frame), is plotted as a function of wavelength (a comparable plot for the NRSIRS2 mode is provided in the instrument manual\footnote{https://jwst-docs.stsci.edu/near-infrared-spectrograph\#NearInfraredSpectrograph-Sensitivityandperformance} in JDox\footnote{https://jwst-docs.stsci.edu/about-jdox}).

In round numbers, in $10\,000$~s NIRSpec is expected to be capable of reaching continuum fluxes lower than $f_\nu\!\simeq\!100$~nJy ($AB\!\simeq\!26.4$) in the lowest resolution $R\!=\!100$ PRISM mode, $f_\nu\!\simeq\!1$~$\mu$Jy ($AB\!\simeq\!23.9$) in $R\!=\!1000$ grating mode and $f_\nu\!\simeq\!3$~$\mu$Jy ($AB\!\simeq\!22.7$) in the highest resolution $R\!=\!2700$ grating mode. It is stressed that these limiting fluxes refer to well-centred point sources. The corresponding sensitivities expressed in terms of the integrated magnitudes of extended galaxies depends on the angular size and detailed surface brightness profile of each object being observed, and its projection on the shutters being employed.

Dealing with arbitrarily shaped extended sources in a realistic manner is beyond the scope of the current MOS data processing and calibration scheme described in Sect.~\ref{subsec:mos_proc}. It assumes that each spectrum either originates from a point source (in which case the calibrated spectrum is radiometrically referenced to the total point source flux at each wavelength) or from a uniform extended source (in which case the calibrated spectrum is referenced to the absolute surface brightness of the region sampled by the shutter on the sky). The NIRSpec MOS pipeline involves a fairly standard sequence of flat-fielding and spectrophotometric correction steps. However, the initial flux calibration assumes as its starting point that all spectra arise from point sources that are perfectly centred in their respective shutters. Deviations from this assumption are subsequently dealt with by applying suitable wavelength-dependent corrective factors to the extracted spectra on a case-by-case basis.

In the uniform surface brightness limit, this correction merely reflects the differences in path losses between a centred point source and a uniform extended source as a function of wavelength (cf. Figs.~11 and 13 of Paper~I), and the solid angle extended by the shutter open area needed to convert the registered signal to a surface brightness. 

In the point source case, an inescapable feature of the MSA is that the majority of targets observed simultaneously in any given MOS exposure cannot be perfectly centred within their respective shutters. In this case the photometric correction needs to reflect the different pathlosses suffered by an off-centred object with respect to a well-centred one. The correction term as a function of the source offset and wavelength is encapsulated in a set of calibration reference files that will be measured during the instrument on-orbit commissioning by stepping a suitable number of targets around within their respective slitlets. 

\begin{figure}
\centering
\includegraphics[width=0.6\hsize]{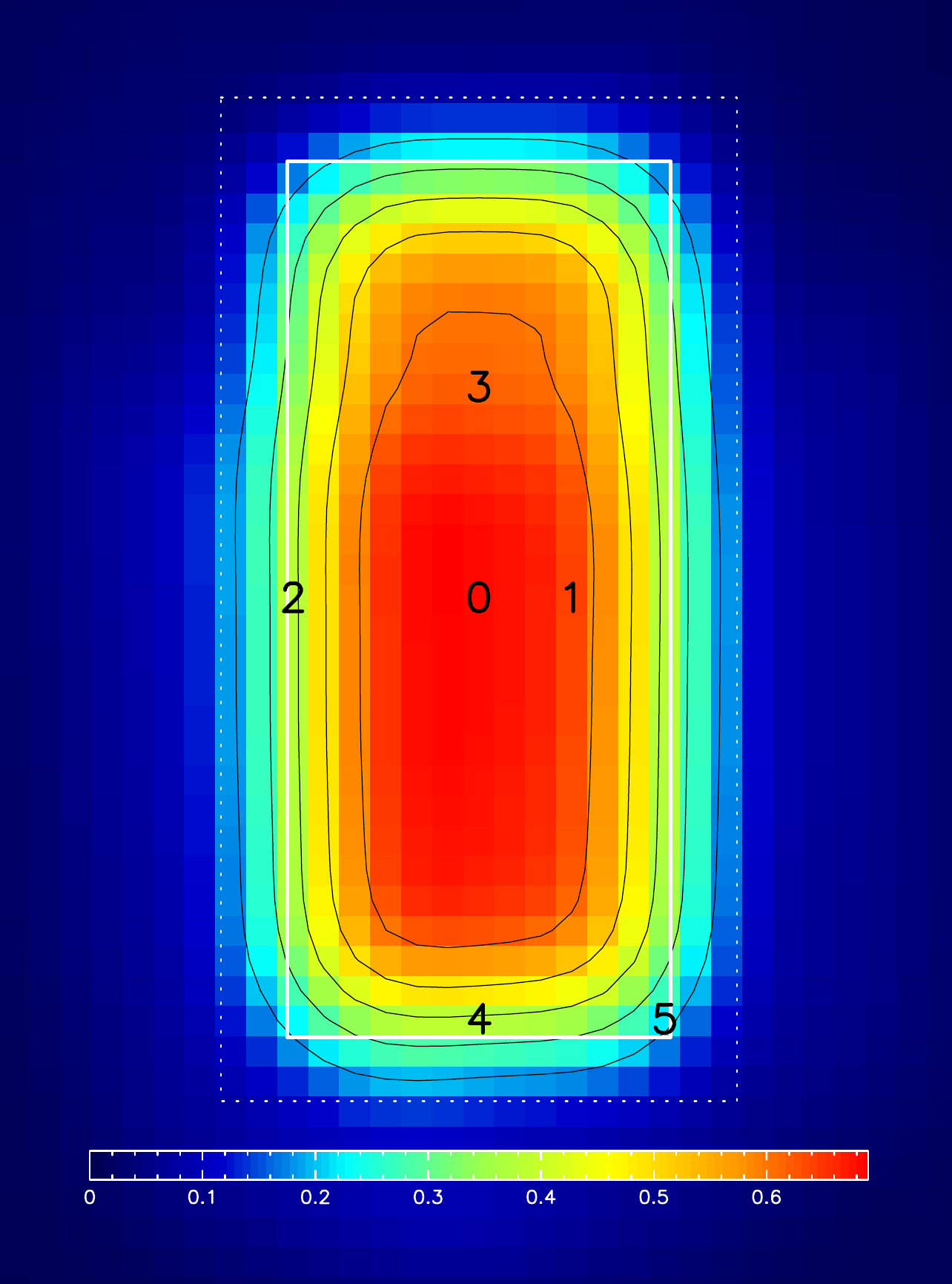}
\caption{Map of the variation in the transmission through a single MSA shutter for a point source at 2.5~$\mu$m as a function of source position. The shutter open area and pitch is marked by the white solid and dotted lines, respectively. The numbers refer to the source locations for which the wavelength dependency of the transmission is shown in Fig.~\ref{fig:pathloss_uni_mos}.}
\label{fig:pathloss_ps_mos}
\end{figure}

\begin{figure}
\centering
\includegraphics[width=\hsize]{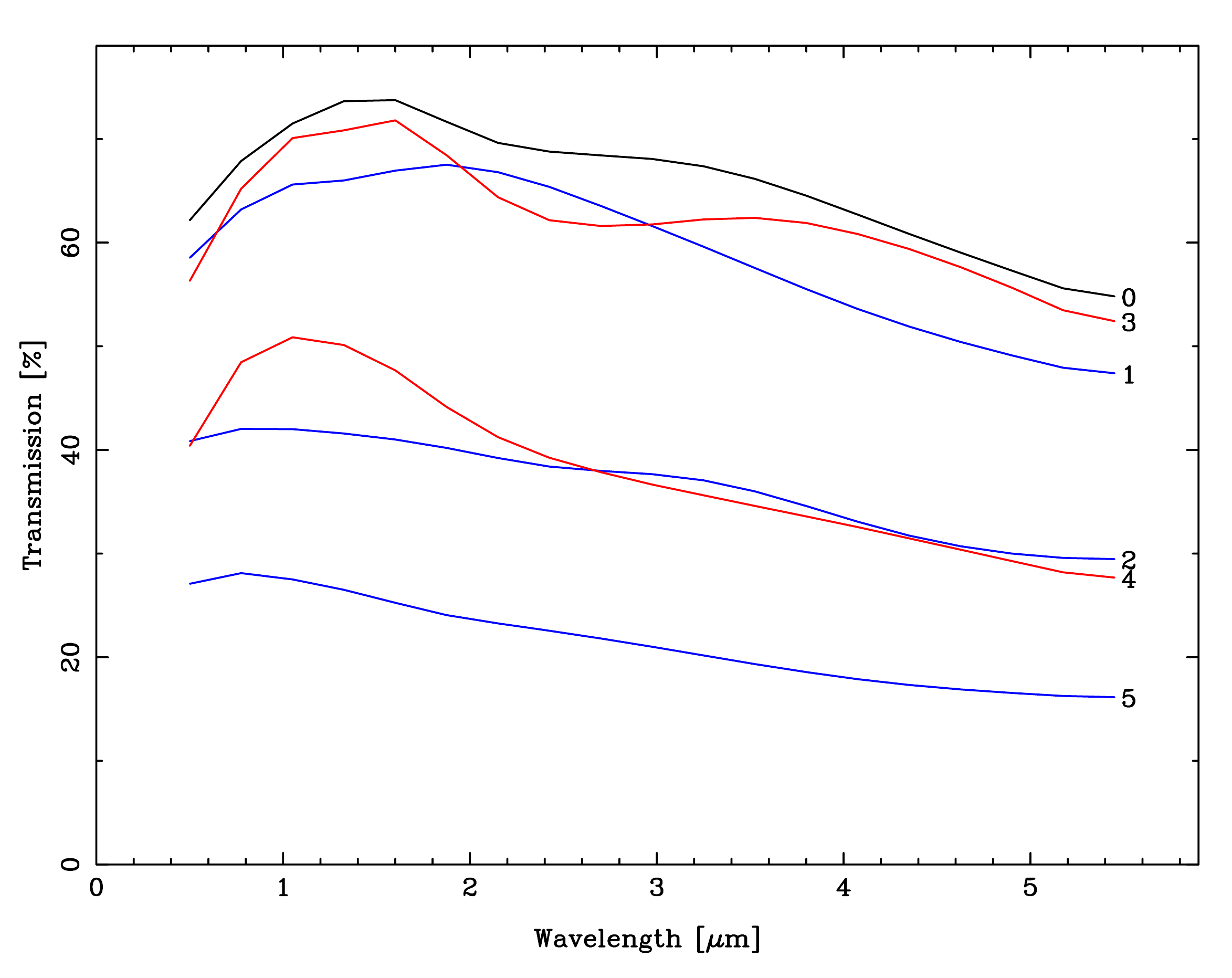}
\caption{Single shutter point source transmission as a function of wavelength for the source locations indicated in Fig.~\ref{fig:pathloss_ps_mos}.}
\label{fig:pathloss_uni_mos}
\end{figure}

The anticipated spatial and wavelength dependency of the transmission through a single MSA shutter is illustrated in Figs.~\ref{fig:pathloss_ps_mos} and \ref{fig:pathloss_uni_mos}, respectively, which result from detailed Fourier optical simulations that include both the geometrical and diffraction components to the total net transmission. Figure~\ref{fig:pathloss_ps_mos} maps a constant wavelength slice through the resulting datacube showing how the total net transmission at $\lambda\!=\!2.5\!~\mu$m varies with target location over a single MSA shutter. Not surprisingly, the net transmission can reach lower values if an Acceptance Zone allowing targets to be placed near or under the inter-shutter bars is employed. Figure~\ref{fig:pathloss_uni_mos} shows the wavelength variation of the single shutter point source transmission for a representative sample of target locations within the shutter. It is apparent that the wavelength dependence of the correction factor is quite pronounced over the face of the shutter open area. 

Figures~\ref{fig:pathloss_ps_mos} and \ref{fig:pathloss_uni_mos} make it clear that accurate knowledge of the relative placement of each point-like target within its slitlet is required if the appropriate spectrophotometric correction is to be applied. A priori information on the anticipated target locations may be had from the MPT or eMPT when designing and setting up the MSA observation. A coarse a posteriori verification can be made provided an undispersed confirmation image is taken following target acquisition (Sect.~\ref{subsec:ta}). However, as is evident from Fig.~\ref{fig:pathloss_ps_mos}, any errors in the assumed target intra-shutter locations translate directly into errors in the spectrophotometric correction factors, especially near the edges of the shutters where the slit transmission drops off the steepest. Thus the choice of Acceptance Zone discussed in the previous section  becomes a trade-off between how many targets can be observed simultaneously, which fraction of their flux is obtained, and how accurately their extracted spectra can be photometrically calibrated. Only actual on-orbit experience will allow the best compromise to be struck between these three objectives.

The computation of the spectrophotometric correction factors of actual, arbitrarily-shaped extended objects that fit neither in the point source category, nor in the uniform source one, will require additional development. If a-priori knowledge of the intrinsic shape of the object is available, it can be combined with the point-source spectrophotometric correction factors by decomposing the object in a collection of point sources with different centrings to derive the net wavelength dependent correction factor applicable to the spatially integrated spectrum of the extended target. Going this extra mile may be very important when, for example, forming diagnostic line ratios from nebular lines observed at different wavelengths. Only very limited information on these spectrophotometric corrections could be obtained during the ground-testing of NIRSpec so, again, actual on-orbit experience will be needed to characterise accurately these effects.

\subsubsection{Impact on wavelength calibration}
\label{sec:wlzp}

If a point source is not perfectly centred within its shutter, this impacts the wavelength calibration of its spectrum. As discussed in Paper~I, the extraction and wavelength calibration of the spectral traces relies on the NIRSpec
instrument model. The model has been tuned to closely match the
as-built flight instrument during the NIRSpec ground calibration
campaign \citep{dorn16, glf+16}, using reference sources (external and
internal to the instrument) that provided a uniform illumination of
the field. Therefore, by default, NIRSpec model-based wavelength
calibration applies to uniform sources and, to first order, to point-like sources centred in their aperture.

\begin{figure}
\centering
\includegraphics[width=0.9\hsize]{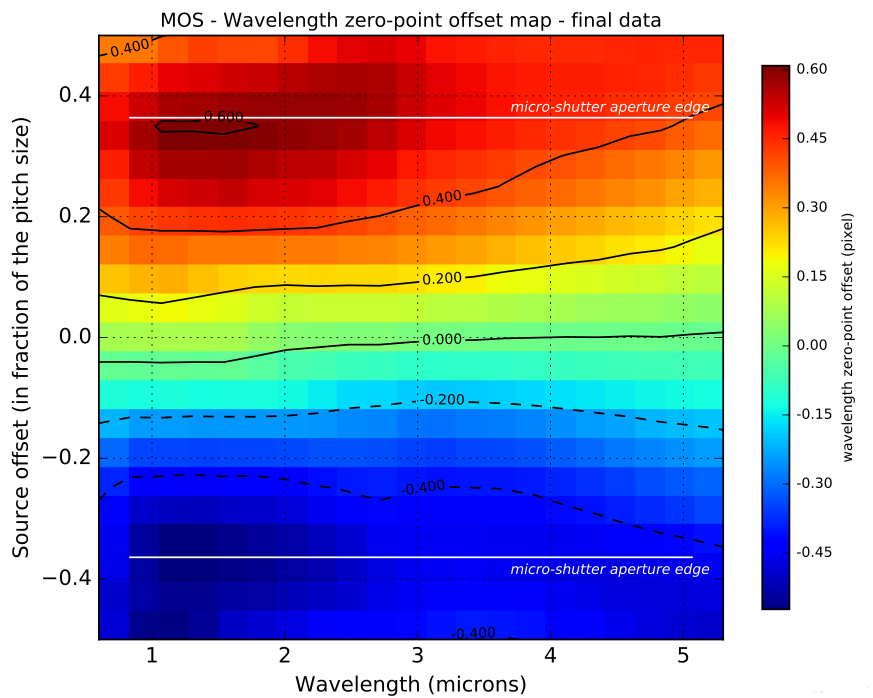}
\caption{Wavelength zero-point offset for a point-source as a function of wavelength and target displacement from the micro-shutter centre in the dispersion direction. For extreme offsets, the core of the point-spread function progressively moves out of the shutter opening and the position of the photocentre becomes sensitive to secondary features like Airy rings. This explains the non-monotonic behaviour of the wavelength zero-point offset at short wavelengths and for large source offsets.}
\label{fig:wlzp_mos}
\end{figure}

For non-centred point sources, the photocentre will in many cases be noticeably offset from the centre of the shutter, in which case an intra-shutter level refinement of the wavelength calibration zero point is applied to the extracted spectrum (Sect.\,\ref{subsec:mos_proc}). Figure~\ref{fig:wlzp_mos} shows a Fourier optical simulation of the difference measured along the dispersion direction between the monochromatic photocentre of an offset point source and that of a uniformly illuminated slitlet as a function of wavelength. As anticipated, the zeropoint correction is most noticeable at the shorter wavelengths where the point spread function is the narrowest.

\begin{figure*}
\centering
\includegraphics[width=0.8\textwidth]{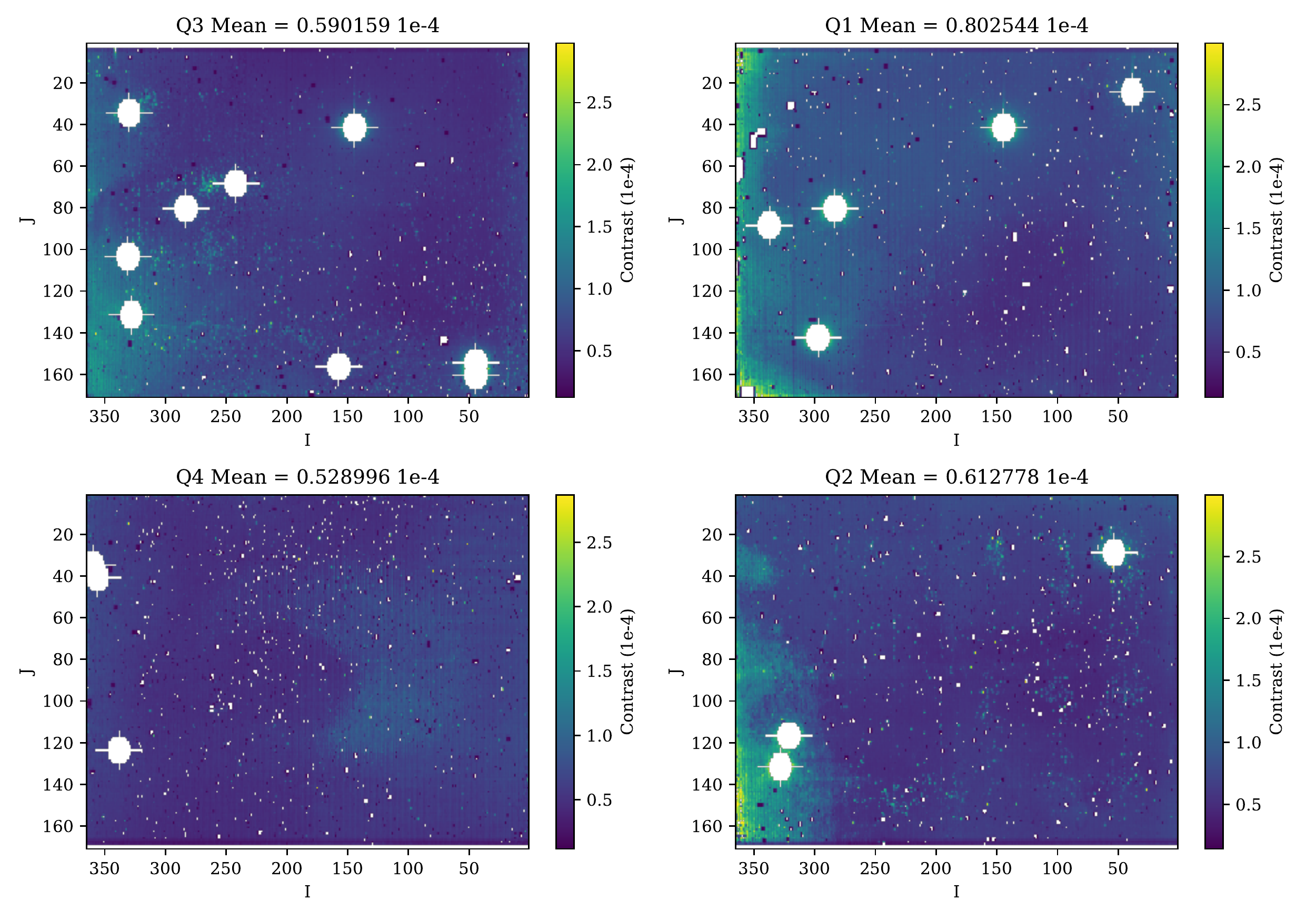}
\caption{Measured attenuation performance of the four MSA quadrants, as derived from internal lamp exposures taken in imaging mode during instrument testing. The axes are labelled according to the MSA plane shutter through which the incident light passed. Pixels in white do not have a measured attenuation due to either flux from heavily saturated failed open shutters or bad pixels on the detector.}
\label{fig:msa_contrast}
\end{figure*}

\subsection{MSA leakage and parasitic background}
\label{sec:contamination}

The closed MSA is not infinitely opaque. It is possible for a small fraction of the incident light to leak around the closed shutter doors and through the array structure itself. A series of dedicated exposures were taken during the ground testing of the instrument to assess the contrast performance of the MSA. Figure~\ref{fig:msa_contrast} shows images of the four MSA quadrants obtained in a long exposure taken with the all-closed MSA illuminated by a bright broadband 0.6-5.3~$\mu$m continuum lamp of the internal calibration source (Paper~I) with the imaging mirror of the grating wheel in place. The undispersed images of the 20 failed open shutters are heavily saturated in this exposure, as are the images of the fixed slits (not shown). The unattenuated incident signal from the lamp was determined by taking separate unsaturated short windowed exposures of the large S1600A1 Fixed Slit. These relative leakage maps (Fig.~\ref{fig:msa_contrast}) show the ratio of the attenuated signal to the inferred incident one. 

The typical contrast level of the MSA is better than one part in $\simeq$5000, equivalent to an attenuation of $\Delta m\simeq9.3$ magnitudes. It is evident that for NIRSpec MOS observations one will need to check that there are no bright objects present in the MSA field of view whose attenuated dispersed images overlap with the spectra of any primary targets. Furthermore, the pile-up of the dispersed images of the entire MSA in the attenuated light of the spatially smooth background sky (cf. Fig.~\ref{fig:detector_gaps}) will give rise to an additional low level sky background signal, which, however, is subtracted out along with the other background components when using the standard nodded slitlet strategy. Since the MOS and IFS modes share the same detector area, observers using the IFU also need to be aware of these sources of parasitic contamination. For the IFS mode, these sources of contamination are enhanced because of the lower throughput with respect to that of the MSA, and the smaller IFU slice width compared to the width of a micro-shutter. This topic is further discussed in Paper~III.

%__________________________________________________________________
\section{Observing in MOS mode}
\label{SectionObservations}

\subsection{Preparing MOS observations}
\label{subsec:prep}

Observers planning to use JWST must generate their proposals within the Astronomer's Proposal Tool (APT)\footnote{http://apt.stsci.edu} from the Space Telescope Science Institute. This application allows the definition of a set of observations through a series of mode-specific templates, controlling all of the configurable instrument and detector parameters, as well as targeting and acquisition, timing constraints, and programmatic choices such as dithering and background strategy. Observers must also upload scientific and technical justifications for review. The fully-defined, template-based observations are directly translatable into on-board scripts for the telescope to execute. Once the Time Allocation Committee (TAC) has selected which proposals will be observed, they are in principle ready for scheduling without further interaction with the observer. However, when it comes to preparing observations in NIRSpec MOS mode, this `single-stream' approach has to be amended.

For targets requiring a particular on-sky orientation of the instrument (e.g. placing a slit along a specific axis of an object) or limited range of orientations (e.g. to avoid having a bright star in the MSA field of view), a special requirement can be added to the observation in APT. The intricacies of the JWST attitude and field-of-regard constraints mean that these orient requirements, at a given location on the sky, translate directly into limitations on the  observation date: for a fixed orientation, the nominal roll flexibility of approximately $\pm$5\,deg corresponds to $\sim$8 day window for observation. Such a constraint increases the scheduling difficulty, and potentially impacts observatory efficiency, so the TAC must adjudicate whether it is scientifically justified, and factors that into their allocation decision.

For a standard MOS programme, where the observer wishes to target as many objects as possible from a catalogue covering a large area on the sky, there is no preferred orientation. However, as discussed in Sect.~\ref{Sec:multi}, MOS observations require very accurate alignment of many science sources within the shutter array, so even in this case the orientation must be known in order to prepare a usable MSA configuration. It would not be efficient to artificially impose an orientation constraint merely to be able to generate an MSA configuration, so instead, a two-step process is employed for MOS mode preparation.

For the initial proposal submission, investigators must demonstrate that they have a workable programmatic strategy, supported by calculations from the JWST Exposure Time Calculator\footnote{https://jwst.etc.stsci.edu/}. Instrument and detector parameters in the APT MOS template must be fully configured, but the submitted MSA configurations need only be placeholders, generated for a real or simulated catalogue at a representative roll angle. For this, investigators can typically use the MPT \citep{kara14}, integrated within APT. Possible orientations for the actual target location can be derived via the JWST General Target Visibility Tool (GTVT)\footnote{https://jwst-docs.stsci.edu/jwst-other-tools/jwst-target-visibility-tools/jwst-general-target-visibility-tool-help}.

\begin{figure*}[]
\includegraphics[width=0.8\textwidth, clip=True]{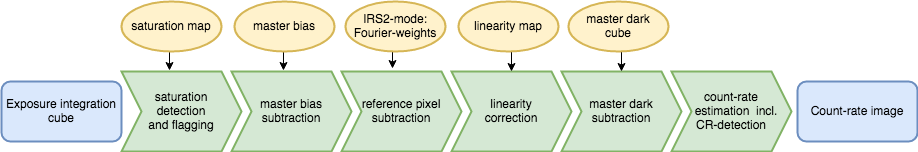}
\\
\includegraphics[width=1.0\textwidth, clip=True]{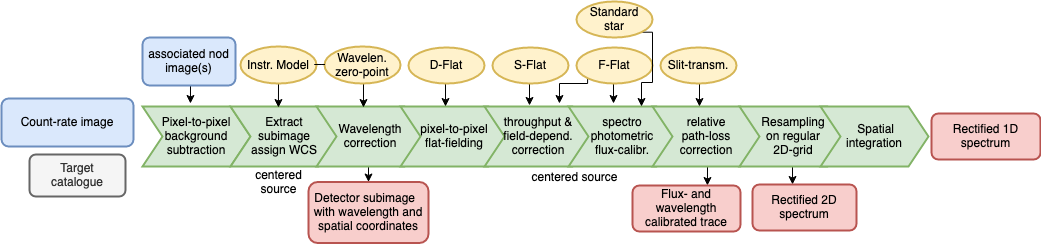}
\caption{{\em Top}: Stage 1 of the NIRSpec processing pipeline: generation
  the count-rate maps starting from the exposure up-the-ramp
  integration cube. {\em Bottom}: Stage 2 of the NIRSpec processing pipeline for a MOS
  observation for point-like sources. The processing is applied to
  each source in the target catalogue that provide the pipeline with
  a source sky position, MOS aperture coordinates $(k, i, j)$ and
  relative coordinates within the slitlet. Steps 2, 5 and 6 are
  valid for an ideally-centred source, steps 3 and 7 
  introduce the necessary corrections taking into account the actual source
  position relative to the slitlet centre. Yellow ovals indicate the reference calibration data necessary to perform the operation.}
\label{fig:proc_steps}
\end{figure*}

Many fields-of-interest lack existing imaging of sufficient breadth or depth, or such imaging is rendered obsolete by proper motion in Galactic fields or shifts between optical and infrared source locations. Therefore, observers have the option to combine NIRCam pre-imaging and NIRSpec MOS observations into the same proposal, to be executed in a single cycle. These NIRCam exposures must be fully configured in the initial proposal submission. Although there are a wide variety of filter options for NIRCam, the F115W filter is particularly well-matched to the NIRSpec F110W filter available for MSA target acquisition (Sect.~\ref{subsec:ta}). Several recommended options also exist to create NIRCam mosaics of sufficient uniformity and coverage for one or more MSA pointings (see JDox for more details on NIRCam pre-imaging\footnote{https://jwst-docs.stsci.edu/near-infrared-spectrograph/nirspec-operations/nirspec-mos-operations/nirspec-mos-operations-pre-imaging-using-nircam}).

After the TAC selection is complete, a long-range plan can be generated, assigning an observation date to each programme, which effectively fixes the orientation angle too. Given the newly assigned orientation, the observer can embark on stage two of the MOS proposal preparation, creating their final flight-ready MSA configurations. Normally, the re-planned MSA configurations are expected to be submitted at least six weeks before the assigned observation execution date\footnote{https://jwst-docs.stsci.edu/display/JDOX/.NIRSpec +MOS+and+MSATA+Observing+Process+v1.1}. For programmes that also requested NIRCam pre-imaging, the second step can only be completed once a catalogue has been generated from the associated imaging. Pre-imaging exposures will be scheduled as early in the Cycle as possible and at least six weeks in advance of the associated MOS execution (\citealt{beck16}; see also the corresponding JDox documentation\footnote{https://jwst-docs.stsci.edu/jwst-near-infrared-spectrograph/nirspec-operations/nirspec-mos-operations/nirspec-mos-and-msata-observing-process}). This means that in some cases very little time will be available to generation the input source catalogues from the NIRCam pre-imaging. Due to the nominal 8 day window for a fixed orient, all flight-ready MSA configurations are only valid for their allocated time slot, so any significant schedule delay will force a re-plan. For current information on MOS mode proposal preparation, consult the latest Call for Proposals and the MOS specific roadmap on JDox\footnote{https://jwst-docs.stsci.edu/methods-and-roadmaps/jwst-multi-object-spectroscopy/mos-roadmap}.

\subsection{Target acquisition} 
\label{subsec:ta}

Most observations using the MOS mode require target acquisition (TA) accuracy of $<$25\,mas, in order to ensure an accurate positioning of the sources in their respective shutters and to limit the spectrophotometric calibration errors associated with the uncertainties in their positioning (Sect.~\ref{sec:slit_tran}). The nominal guide-star acquisition pointing accuracy for JWST is predicted to be $\sim$100\,mas ($1\sigma$ error, per axis\footnote{https://jwst-docs.stsci.edu/jwst-observatory-characteristics/jwst-pointing-performanc}). Therefore, the default for MOS mode observations will be to start with a MSA Target Acquisition (MSATA) which, optimally, can increase the accuracy to $\sim$20\,mas. Skipping the MSATA is allowed in the MOS mode APT template, but is only recommended if the science region-of-interest is sufficiently extended that exact placement is unwarranted.

During the automated MSATA procedure, a set of $\sim$5--8 TA reference targets (typically stars, although suitably bright and compact galaxies may also be used) are observed in imaging mode through the MSA, in either an `all open' configuration or a bespoke `protected' configuration if it is necessary to block out bright objects in the field. Two exposures separated by an MSA half-shutter slew ensure that all the reference targets are detected, regardless of the initial guide-star acquisition accuracy. The on-board software measures the centroid locations of the reference objects, determining their mean offset and hence the corrective slew required by the observatory before beginning the primary MOS observation \citep{keye18}.

The expected accuracy of MSATA depends significantly on the number of available reference targets in the field, their surface brightness profile central concentration, and (importantly) the relative astrometry of the reference objects to the science targets. Optimal MSATA accuracy requires that all reference {\em and} science objects are in a single catalogue possessing a relative astrometric accuracy of $\sim$5\,mas. Source catalogues derived from images acquired by ACS or WFC3 on the Hubble Space Telescope generally meet this requirement, as will NIRCam imaging data.

\subsection{Data processing and calibration}
\label{subsec:mos_proc}

Figure~\ref{fig:proc_steps}
illustrates the main data processing and calibration steps \citep{alve18} that have to
be applied to MOS raw data (level 1a, i.e. the up-the-ramp integration-cubes) to
derive flux and wavelength calibrated spectra for each individual
target. This workflow and the associated algorithms are used both by the NIRSpec instrument team data processing software and the STScI science calibration pipeline\footnote{https://jwst-docs.stsci.edu/jwst-science-calibration-pipeline-overview}.
We note that Fig.~\ref{fig:proc_steps} only includes the steps corresponding to Stage~1 and Stage~2\footnote{https://jwst-docs.stsci.edu/jwst-science-calibration-pipeline-overview/stages-of-jwst-data-processing} of the data processing. A description of Stage~3 that combines fully calibrated data from multiple exposures is beyond the scope of this paper.

Like for all the other NIRSpec modes, and indeed for all the raw data
from JWST science instruments, the first stage of the data reduction
process generates the up-the-ramp slope or count-rate images from the raw data
cubes. Because this data-reduction stage is not specific to the instrument mode, it is referred to as the pre-processing or Stage-1 pipeline. This stage involves the following steps: saturation
detection and flagging, master bias subtraction, reference pixel
subtraction, linearity correction, dark subtraction,\footnote{The dark
  subtraction is carried out at the data cube level, i.e. the
  corresponding frame from a low-noise dark-currents cube is
  subtracted from each frame of the exposure.}  and count rate
estimation (including jump detection and cosmic ray rejection) -- see also
\cite{NTN-2011-004}, \cite{bbm+12} and \cite{giar19}.

The second stage of the data processing concerns the extraction of
calibrated spectra from the NIRSpec count-rate images. Together with the count-rate images, the other main
input to this stage is the observed target catalogue, where for each target
the following basic information has to be listed: the target's
sky-coordinates; the target's position in the MSA array in terms of its shutter
ID $(k, i, j)$ and the relative XY-coordinates within the shutter
pitch; whether the source is point-like or extended; and finally
the background subtraction strategy to be used for the target (available options: none,
use all the flanking shutters in the nodded 1x3 slitlet; use only some of these; or use a
master background).

%The default processing steps to extract the wavelength-calibrated spectra from
%the count-rate images for a point source are then the following:
%
%\begin{itemize}
%
%\item[1] Pixel-level background subtraction (combining the
%exposures from the three-nods) if relevant
%
%\item[2] Extraction of the sub-image containing the spectral-trace and assignment of
%wavelength and spatial coordinates to each pixel therein
%
%\item[3] Wavelength zero-point correction (see Sect.\,\ref{sec:wlzp})
%
%\item[4] Pixel-to-pixel flat-field correction (D-flat)
%
%\item[5] Spectral flat fielding: correcting the count rate for field-
%  and wavelength-dependent variations in the throughput such as the
%  grating blaze function (relative flux calibration) (S-flat and F-flat)
%
%\item[6] Conversion to absolute flux-units by comparing the
%  count-rates in the subimage, to those from a spectro-photometric
%  standard star (absolute flux calibration)
%
%\item[7] Correction for path-losses relative to a centred source
%
%\end{itemize}

The seven default processing steps needed to extract wavelength-calibrated point source spectre from the count-rate images are: {1)} The pixel-level background subtraction is performed (combining the exposures from the three-nods) if relevant. {2)} The sub-image containing the spectral trace is extracted and wavelength and spatial coordinates are assigned to each pixel therein. {3)} The  wavelength zero-point correction (Sect.\,\ref{sec:wlzp}) is applied. {4)} The pixel-to-pixel flat-field correction is performed (D-flat). {5)} Spectral flat fielding, that is, correcting the count rates in the subimage for field- and wavelength-dependent variations in the throughput such as the
grating blaze function (relative flux calibration) (S-flat and F-flat). {6)} Conversion to absolute flux-units by comparing the
count-rates in the to those from a spectro-photometric
standard star (absolute flux calibration). {7)} Applying the
correction for path-losses relative to a perfectly centred source.

In the case of a spatially-uniform source, step 3 is omitted, and an
additional `pixel-area' correction is applied to convert data from 
flux-per-pixel to surface brightness units. For a uniform
source, the path loss correction is applied relative to a centred
point-source and additionally combined with a correction for the shadow cast by the opaque bars present between the shutters of a slitlet.

If the source is marked as extended, the standard pipeline will treat it like a spatially-uniform source. One will therefore have to derive
additional, object-specific correction factors to be applied to standard pipeline data products to supplement the missing or approximate correction factors applied as part of steps 3 and 7. As highlighted in Sect.\ref{sec:slit_tran}, generating an object-specific correction for step 7 may require a-priori knowledge of the intrinsic shape of the object at the wavelengths of interest.

If the options to not correct for background or to use a master-background have been specified for the target,
then step 1 is skipped. The subtraction of a
master-background spectrum (derived for example by combining the
background data gathered through the fixed-slits and/or individual
open shutters containing empty sky) from the target data is applied to the
target's subimage after the wavelength-coordinates have been assigned
to the pixels in step 2. The final, exposure-level data products for each target are: i) flux and detector-subimage containing the spectral trace with
wavelength and spatial-coordinates assigned, ii) the spectrum
re-sampled on a rectified 2D-grid; and iii) the 1D-spectrum obtained by spatially
collapsing the data in the 2D-grid. In the following section these processing and calibration steps are applied to a simulated NIRSpec data set.

\begin{table*}
\caption{Breakdown of the sources included in the simulation. The last two columns provide the number of sources in each priority class contained in the input mock catalogue and the number of such objects selected (in any one of the three dithers). The galaxy properties (redshifts, magnitudes and line intensities) used to define the priority classes are all derived from the simulated NIRCam photometric information present in the mock input catalogue. In particular the redshifts and the line intensities were obtained by fitting the spectral energy distribution of the objects derived from the mock photometric points. We note that the definition of the priority classes and in particular of the various thresholds that were used depends on the depth of the observations as well as on the scientific goals of the survey. Therefore, they may not be applicable directly to other observations.} 
\label{tab:targets} 
\centering
\begin{tabular}{@{}lccc}
\hline
Galaxy properties & Priority class & Mock catalogue & Selected \\
\hline\hline
\noalign{\smallskip}
Place-holder for rare sources (not used)  &  1                    &   0          & 0 \\
$z \geq 10$, $AB < 29.5$ &   2            & 17       & 9\\
$z \geq 10$, $29.5 < m_{\rm F200W} < 30.5 $  & 3  & 31  & 4 \\
$z>6$, S/N$_{\mathrm{H}{\alpha}} > 5$, S/N$_{\mathrm{H}{\beta}} > 8$ & 4 & 100 & 7\\
$z>2$, S/N$_{\rm continuum} > 30$ &   5 &       139        & 13\\
$z>6$, S/N $<$ class 4 &   6 &       395               &  41\\
$1.5<z<6.5$, $27.25<m_{\rm F444W}<28.50$ &   7 &  4404    & 160\\
$m_{\rm F444W}<=29$ &   8 &     14411                     & 105 \\
Fillers   & 100 &   33319                               & 31\\
Total &   &   52817        & 370 \\
\hline
\end{tabular}
\end{table*}

\begin{figure*} 
\centering
\includegraphics[width=0.8\textwidth]{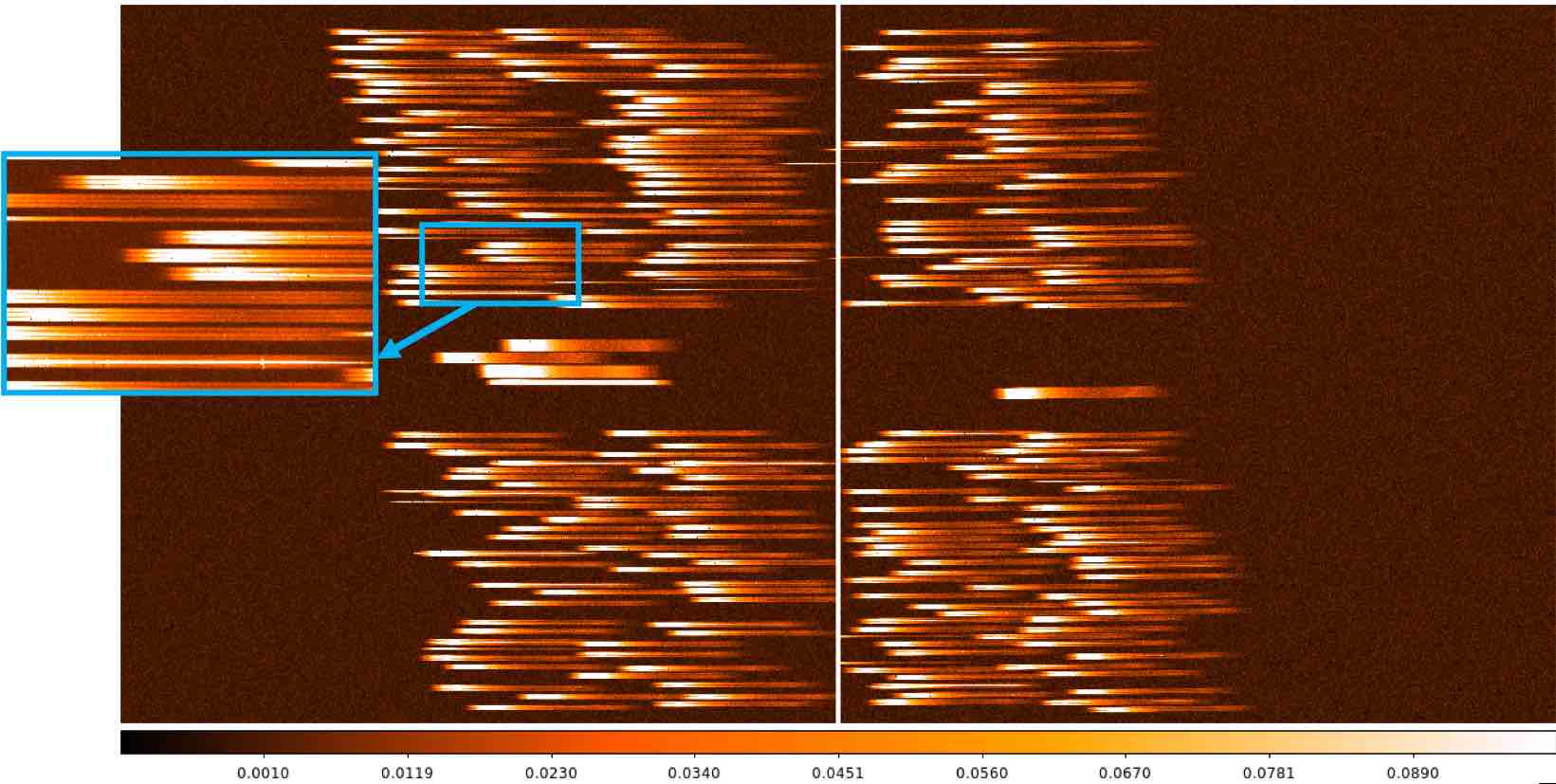}
\caption{Simulated count-rate image of one of the MOS-mode PRISM exposures. There are 214 targets covered in this MSA configuration, each one observed within a 3$\times$1 shutter slitlet. The noise level corresponds to that of a single exposure of 2801\!~\!s duration. The inset provides a close-up of a few representative traces. The spectra of several of the stronger sources are visible above the sky background, as is the spurious single-shutter sky spectrum from a defective FO shutter (fourth spectrum from the top in the inset).}
\label{fig:prism_d0}
\end{figure*} 

\begin{figure*}
\centering
\includegraphics[width=0.8\textwidth]{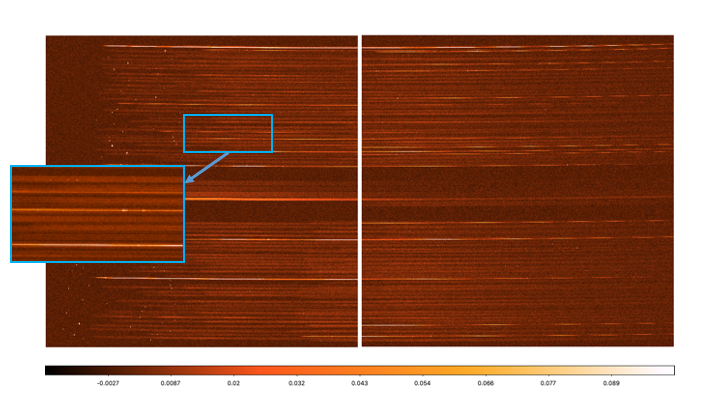}
\caption{Simulated count-rate image for the G235M grating exposure matching the PRISM exposure of Fig.~\ref{fig:prism_d0} and employing the same MSA configuration. The noise level corresponds to that of a single exposure of 2801\!~\!s duration. One of the spectra at the middle of the inset shows examples of emission lines.}
\label{fig:g235m_d0}
\end{figure*}

\section{Simulated MOS observations: a spectroscopic deep field}
\label{SectionSimulation}

The functioning of NIRSpec in MOS mode is well illustrated
by a set of simulations designed to mimic the deep spectroscopic
observations that will be performed as part of the JWST Advanced Deep
Extragalactic Survey (JADES), a joint project of the NIRCam and
NIRSpec Guaranteed Time Observation (GTO) teams. The simulation aims to reproduce the JADES deep spectroscopic follow-up of high-redshift sources in the GOODS-South field identified through prior deep imaging observations with NIRCam that can be found in GTO programme 1287\footnote{https://jwst.stsci.edu/observing-programs/program-information}). The simulated programme consists of a 92~ks MOS observation in the $R\!=\!100$ CLEAR/PRISM (broken up into three dithered pointings, each with its own MSA configuration), combined with matching 3$\times$25~ks observations in each of the F070LP/G140M, F170LP/G235M, F290LP/G395M $R\!=\!1000$ gratings. The observations being simulated have all targets observed in three shutter tall slitlets, with the targets nodded between the shutters of the slitlet between sub-exposures. Most notably, it has the higher resolution $R\!=\!1000$ grating observations being carried out using the MSA configurations and pointings optimised for the PRISM observations. This results in overlapping of the grating spectra (Sect.~\ref{SectionPerformances}), whose emphasis, however, is on measuring emission lines. The objective is to derive the source continuum spectra from the
PRISM observations, together with the strongest lines, and then use this information to disentangle the overlapping grating spectra\footnote{This step is not part of the standard NIRSpec data processing pipeline.}. This approach is adopted to enable the same set of targets to be observed with both the PRISM and the gratings.  

The input catalogue used for the simulation consisted of 52\,817 targets contained within a circular 3~arcmin radius area extracted from the 
mock catalogue of the JAdes extraGalactic Ultradeep Artificial
Realizations (JAGUAR) package, developed by \cite{wch+2018}. The input catalogue targets were sorted into 8 priority classes, plus fillers, as
summarised in Table~\ref{tab:targets}. The eMPT software suite \citep{pbs+2020} was used to place the targets in each priority class in sequence on the MSA (Sect.~\ref{Sec:multi}), and the three dithered pointings were chosen to maximise the coverage of the highest priority targets and the number of common targets (i.e. observed in all three pointings). 

Over the three different dithered pointings (each of them corresponding to a set of three nodding positions), the eMPT algorithms selected a total of 370 unique targets. To maximise the number of objects in common to the three pointings, these dithers were only separated by a small distance from each other, that is typically no more than the size of a handful of shutters. Despite that, only 23\% of these targets were covered in all three pointings and 51\% were covered in a single exposure. The limited level of target commonality between the three dithered pointings is mainly caused by the failed closed shutters scattered across the arrays which make the autocorrelation function of the pattern of operational 3-shutter slitlets drop off steeply even at small displacements.

Detailed simulations of each of the three nodded exposures for each of the three dithered pointings in each of the four utilised dispersers were then produced using the template spectra from the JAGUAR input catalogue, treating each target as a point source for computational reasons. The simulations also included the sky background emission, which is comprised of the in-field zodiacal light, the parasitic light, and thermal emission from the telescope \citep{light16}. The same breakdown of the individual exposures into sub-exposures as per the GTO programme was followed, and the detector noise model of \cite{bsf+2018} was assumed.

\begin{figure}
\resizebox{\hsize}{!}{\includegraphics{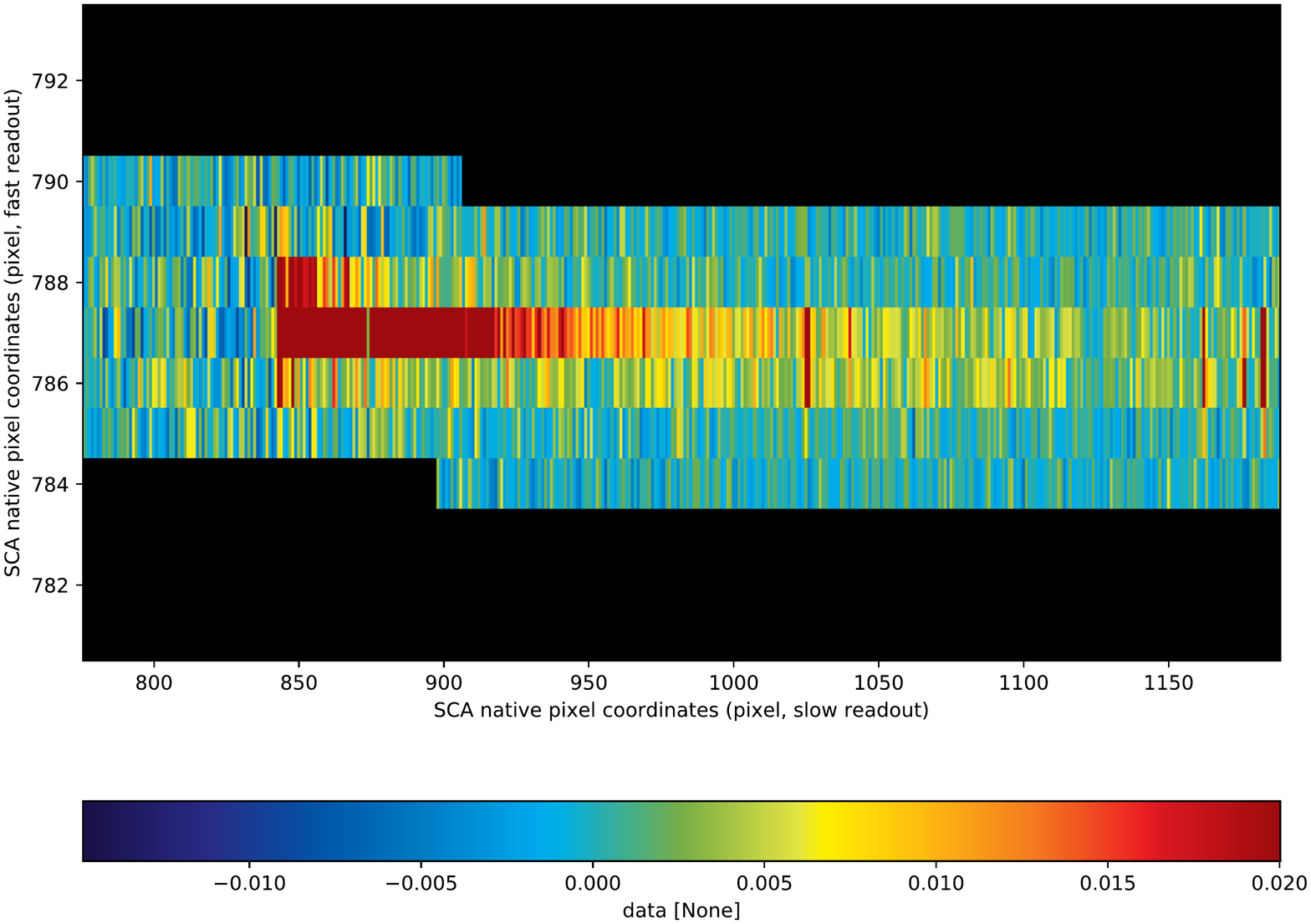}}
\resizebox{\hsize}{!}{\includegraphics[width=\hsize]{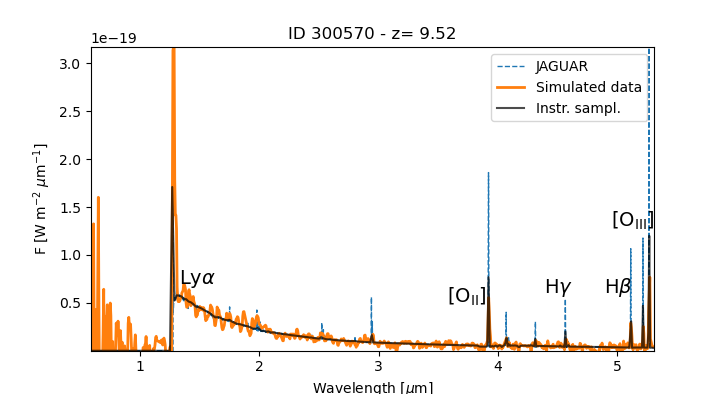}}
\caption{Example of a simulated NIRSpec MOS spectrum. {\it Top panel:} Sub-image containing the spectral trace extracted from one of the simulated count-rate images in PRISM mode of a mock galaxy at $z\!=\!9.52$ having a mass of $M\!=\!2.9\times10^8~M_{\odot}$, a UV luminosity of $\sim3.6\times10^{28}~{\rm erg~s^{-1}}$ and an $AB$ magnitude of 27.6. {\it Bottom Panel:} The matching final reduced 1D spectrum plotted in orange. The blue line shows the input spectrum from JAGUAR, and the grey line the profile expected from binning the input spectrum to the detector sampling.}
\label{fig:ex_prism}
\end{figure}

The simulated count-rate image for the central nod of one of the PRISM
exposures is shown in Fig.~\ref{fig:prism_d0}, and that of the matching 
G235M grating exposure is shown in Fig.~\ref{fig:g235m_d0}. The spectral overlap occurring in the grating exposure is evident. The noise level in both examples corresponds to the co-addition of two NRSIRS2 integrations spanning 19 groups each, for a total exposure duration of 2801~s. A total of 214 spectra are obtained for this MSA configuration (209 and 214 in the two other dithers), compared to the 232 spectra predicted by Eqs.~(1) through (6) of Sect.~\ref{Sec:multi} for the input target catalogue density ($\Sigma_\mathrm{T}=1870$~arcmin$^{-2}$) and Acceptance Zone ($\theta_x\!=\!0.343$, $\theta_y\!=\!0.420$) employed. This slight reduction in multiplexing compared to the maximum achievable is the statistical price paid for sorting the targets into priority classes.

Individual spectra of the targets were extracted from the simulated count-rate images
using the NIRSpec pipeline described in Sect.~\ref{subsec:mos_proc}. Figure~\ref{fig:ex_prism} shows an example of the extracted background-subtracted spectral trace of a galaxy at $z=9.52$. In the lower panel the final 1D-spectrum is
compared to the assumed JAGUAR input catalogue spectrum. Note the clear detection of the  Ly$\alpha$, $[$O{\sc ii}$]$, $[$O{\sc iii}$]$, H${\beta}$ and H${\gamma}$ emission lines and of part of the galaxy continuum. The count-rate images and the extracted products for all the targets of the above simulation are publicly available from the ESA
website\footnote{www.cosmos.esa.int/web/jwst-nirspec-simulations}.

%__________________________________________________________________

%__________________________________________________________________
\section{Conclusions}
\label{SectionConclusion}

In this paper we have described the multi-object capabilities of NIRSpec employing its novel MSA. We have also outlined the baseline operational and data reduction approaches that have been developed pre-launch to address the various unique issues presented by the device.

NIRSpec is a complex instrument, and especially so in MOS mode. Consequently, both the software tools necessary to plan an observation employing the MSA, and the foreseen pipeline needed to reduce the data that it produces, rely entirely on the availability of the high fidelity instrument model \citep{dorn16,birk16,glf+16} that is used to predict the dispersion and traces of the spectra produced by each shutter. This model will be verified and updated during the on-orbit instrument commissioning programme. This approach of employing the instrument model as the backbone in the planning and reduction of NIRSpec observations is not likely to change in the future.

The baseline operational scenario described in this paper, which assumes that bad pixels and other defects on the NIRSpec detector arrays make it necessary to frequently shift the pattern of dispersed spectra projected on the detector arrays, is a cautious one. It assumes that all targets have to be observed in three shutter tall slitlets, with the telescope pointing sequentially nodded so as to shift the targets between the three slitlet shutters between sub-exposures. In addition, longer duration exposures are assumed to be broken up into several dithered sub-exposures employing slightly offset pointings and different MSA configurations. The statistics of the MSA utilisation presented in Sect.~\ref{Sec:multi}, show that assigning three operational shutters to each target comes at a price, in that the $\simeq14$\% failed and non-operational shutters presently carried by the MSA render it only $\simeq$55\% efficient compared to a fully functional device. moreover, only a few percent of the targets contained in an oversized input catalogue can be expected to be covered in a single MSA exposure.

Despite these limitations, the expected multiplexing at $R\!\simeq\!100$ and $R\!\simeq\!1000$ in fields with a high density of objects exceeds 200 and 50, respectively. It is possible that the master background subtraction approach will eventually be developed to a state where assigning two-shutter or even single-shutter slitlets to each observed target will be attempted in order to achieve higher levels of multiplexing when, for example, carrying out future NIRSpec follow-up observations of deep NIRCam surveys.

As we have seen in Sects.~\ref{sec:slit_tran} and \ref{subsec:mos_proc}, carrying out spectroscopy over roughly three octaves in wavelength with a fixed-sized slit, illuminated by a PSF whose footprint increases as the square of the wavelength beyond 2~$\mu$m where it is diffraction limited, presents its own particular set of challenges. We have considered the two opposing extreme cases where the observed object is either a point source or a large extended source of uniform surface brightness. The processing and calibration of MOS spectra for these two categories of objects do not require any additional a priori knowledge on the shape of the object. In the intermediate cases of actual complex-shaped galaxies, the NIRSpec `beam' at each wavelength formed by the convolution of the top-hat shutter open area and the varying-sized PSF leads to the resulting NIRSpec spectrum effectively probing a significantly smaller weighted segment of the galaxy at its blue end compared to its red end. Only actual on-orbit experience will reveal whether this effect complicates vital tasks such as forming accurate diagnostic line ratios from nebular lines measured in spectra obtained in different NIRSpec grating bands. Despite these challenges and remaining uncertainties -- like the evolution of the pattern of operational micro-shutters during launch and with use in orbit -- NIRSpec's predicted in-orbit MOS sensitivity (Fig.~\ref{fig:mos_sensitivity}) and multiplex capabilities (Sect.~\ref{Sec:multi}) are unprecedented, confirming that JWST, and NIRSpec in particular, will bring near-infrared spectroscopy one step further.

%__________________________________________________________________
\begin{acknowledgements}
NIRSpec owes its existence to the dedication and years of hard work of a great number of engineers, scientists and managers scattered across European and US industry, academia, and the ESA and NASA science programmes. The contributions of these colleagues and institutions are too numerous to list here, but are greatly appreciated all the same. We would like to thank  Nina Bonaventura,  Jacopo Chevallard and Emma Curtis Lake for their contributions in performing the NIRSpec simulations.

\end{acknowledgements}
%__________________________________________________________________
\bibliographystyle{aa}
\bibliography{thebibliography}
%__________________________________________
\end{document}